\documentclass[3p, authoryear]{elsarticle}
\pdfoutput=1
\usepackage{natbib}
\usepackage{enumitem}
\usepackage{url}
\usepackage{booktabs}
\usepackage{array}
\usepackage{xcolor}
\usepackage{rotating}
\usepackage{pifont}
\usepackage{amsfonts}
\usepackage{amssymb}
\usepackage{xfrac}
\usepackage{mathtools}
\usepackage{bm} 
\usepackage{adjustbox} 
\usepackage{tikz}
\usepackage{csquotes}
\usepackage{commath}
\usepackage{graphicx}
\usepackage[labelformat=simple]{subcaption}
\usepackage[]{algorithm}
\usepackage{algorithmic}
\usepackage{float}
\usepackage{placeins} 
\usepackage{pdflscape}
\usepackage{hyperref}
\hypersetup
{
  colorlinks=true,
  linkcolor=blue,
  filecolor=red,
  urlcolor=magenta,
  breaklinks=true
}

\providecommand{\erdos}[1]{Erd\H{o}s-R\'enyi #1}
\providecommand{\ba}[1]{Barab{\'a}si-Albert #1}

\providecommand{\singlehalf}[1]{\textcolor{black!50}{[\textbf{1.5 COLUMN}]} #1}
\providecommand{\double}[1]{\textcolor{black!50}{[\textbf{DOUBLE COLUMN}]} #1}
\providecommand{\incolour}[1]{\textcolor{black!50}{[\textbf{IN COLOUR}]} #1}

\renewcommand{\today}{%
  \ifcase\day\or
  1st\or 2nd\or 3rd\or 4th\or 5th\or
  6th\or 7th\or 8th\or 9th\or 10th\or
  11th\or 12th\or 13th\or 14th\or 15th\or
  16th\or 17th\or 18th\or 19th\or 20th\or
  21st\or 22nd\or 23rd\or 24th\or 25th\or
  26th\or 27th\or 28th\or 29th\or 30th\or
  31st\fi~\ifcase\month\or January\or February\or March\or
  April\or May\or June\or July\or August\or September\or October\or
  November\or December\fi \space \number\year%
}

\usetikzlibrary{arrows, shapes, positioning, plotmarks}
\tikzset{
  >=stealth',
  pil/.style={
    ->,
    thick
    shorten <=2pt,
    shorten >=2pt,
  }
}

\journal{J.}

\begin{document}

\begin{frontmatter}

\title{Understanding Flash Crash Contagion and Systemic Risk:\\A Micro-Macro Agent-Based Approach}

\author[oxcs,oxam]{James Paulin\corref{corr}}
\ead{james.paulin@cs.ox.ac.uk}

\author[oxcs]{Anisoara Calinescu}
\ead{ani.calinescu@cs.ox.ac.uk}
\author[oxcs]{Michael Wooldridge}
\ead{mjw@cs.ox.ac.uk}

\cortext[corr]{Corresponding author}
\address[oxcs]{Department of Computer Science, University of Oxford, Oxford, OX1 3QD, UK.}
\address[oxam]{OxFORD Asset Management, OxAM House, 6 George Street, Oxford, OX1 2BW, UK.}

\begin{abstract}

The purpose of this paper is to advance the understanding of the conditions that give rise to flash crash contagion, particularly with respect to overlapping asset portfolio crowding. To this end,
we designed, implemented, and assessed a hybrid micro-macro agent-based model, where price impact arises endogenously through the limit order placement activity of algorithmic traders. Our novel hybrid microscopic and macroscopic model allows us to quantify systemic risk not just in terms of system stability, but also in terms of the speed of financial distress propagation over intraday timescales. We find that systemic risk is strongly dependent on the behaviour of algorithmic traders, on leverage management practices, and on network topology. Our results demonstrate that, for high-crowding regimes, contagion speed is a non-monotone function of portfolio diversification. We also find the surprising result that, in certain circumstances, increased portfolio crowding is \emph{beneficial} to systemic stability. We are not aware of previous studies that have exhibited this phenomenon, and our results establish the importance of considering non-uniform asset allocations in future studies. Finally, we characterise the time window available for regulatory interventions during the propagation of flash crash distress, with results suggesting \emph{ex ante} precautions may have higher efficacy than \emph{ex post} reactions.

\end{abstract}


\end{frontmatter}

\section{Introduction}
\label{sec:introduction}

During the \emph{Flash Crash} of 6th May 2010, over a trillion dollars were wiped off the value of US companies in an event that has been largely attributed to the rise of \emph{algorithmic trading} \citep{kirilenko2017flash}. Although the circumstances giving rise to the Flash Crash have been intensely studied since 2010, the role played by algorithmic traders in propagating systemic risk through financial networks remains poorly understood. In the context of networks representing the overlap of asset portfolios held by investment funds (hereafter \emph{funds}), systemic risk arises as a result of price impact \citep{brunnermeier2009market}. Despite this, over-simplified systemic risk models typically assume non-crisis linear price impact functions \citep{benoit2016risks}. Systemic events such as the Flash Crash of 6th May 2010 demonstrate that during crises (conditions of abnormal perturbation when system dynamics are far from equilibrium), price impact can deviate strongly from linearity \citep{kirilenko2017flash}. Analyses of flash crashes have found that the market's ability to facilitate price discovery is strongly dependent on the micro-level behaviour of market participants, in particular those involved in algorithmic trading \citep{cftc2010findings, menkveld2016economics}.

In the present work, we simulate crisis price impact due to algorithmic trading with a limit order book (LOB) agent-based model (ABM), calibrated to exhibit extreme transient price impact phenomena such as flash crashes. The micro-model is derived from \citet{paddrik2012agent}. We combine this micro-level ABM with a macro-level network model of overlapping leveraged asset portfolios held by funds. Our paper presents a novel synthesis of models from both the microstructure and the macroeconomic literatures. We are aware of few related studies that attempt such a synthesis \citep{gerig2015high,torii2015shock}. We investigate a key interface between the two classes of model by addressing the over-simplified price impact assumptions typical of macroscopic systemic risk models. In reality, asset prices are formed due to the action of individual traders placing limit orders on stock exchanges, and this is true whether the trader is a global bank seeking to reduce position sizes over several days, or a \emph{high-frequency trader} \citep[HFT,][]{menkveld2016economics} targeting fleeting profit opportunities lasting fractions of a second.
However, assets do not exist in isolation. Rather, they are held within the portfolios of funds. Funds are thus connected to each other by virtue of the assets they hold in common. Similarly, assets are connected to each other by virtue of being held in the same portfolio at a fund \citep{allen2008networks}. Following established methodology from the macroscopic systemic risk literature \citep{huang2013cascading, caccioli2014stability, cont2017fire}, we represent the \emph{fund-asset network} as a bipartite graph. We extend previous works by controlling network topology with a \emph{preferential attachment} method \citep{barabasi1999preferential} which allows us to tune the degree of portfolio overlap and hence the connectivity between funds.

When funds borrow from a bank or brokerage to finance an asset portfolio, they are required to maintain leverage (the ratio of portfolio value to collateral) below a pre-agreed limit as the value of the assets in their portfolio varies \citep{brunnermeier2009market}. If the portfolio value suffers a loss due to price fluctuation, the fund may be required to either deposit additional capital or rapidly liquidate assets (\emph{deleverage}) in order to satisfy the leverage constraint. This scenario is known as a \emph{margin call} and can be enacted by the lending bank on a daily or intraday (during trading hours) basis \citep{gai2011complexity, brunnermeier2009market}. Loss to the lender bank accrues when the liquidation value of a client fund's portfolio is less than the value of the outstanding loans for which the portfolio acts as collateral. In this circumstance, it is impossible for the fund to repay its debt, and the fund \emph{defaults}. As \citet{brunnermeier2009market} describe, a problematic amplification mechanism known as a \emph{margin spiral} can occur if a fund is forced to sell its assets due to losses: the action of selling further depresses prices, increasing losses and requires further rounds of \emph{distressed selling}. Systemic risk arises when different funds have investments in the same assets. Here, distressed selling by a fund causing prices to fall via price impact will cause losses at all other funds with investments in the distressed assets. These other funds may then be required to reduce the size of their investments in order to manage their leverage and so the distress propagates in what is known as a \emph{fire sale} \citep{shleifer2011fire}.

Our hybrid micro-macro ABM allows us to characterise the effect on systemic risk of \emph{leverage management} policies and fund-asset network topology. In addition to considering systemic stability (characterised by the default of funds on their margin loans), the use of a limit order book ABM allows us to measure the \emph{speed} at which financial distress propagates between assets over short intraday timescales characteristic of flash crashes. We investigate systemic risk due to (i) portfolio \emph{crowding}, which is a measure of the similarity of the asset portfolios held by different funds, and due to (ii) portfolio \emph{diversification} which relates to the number of assets in which funds invest. We further investigate the effects on systemic risk of the amount of leverage and capital deployed by funds, and of the tolerance with which funds manage their leverage under fluctuating asset price conditions. We show how these parameters control liquidity supply due to distressed selling in the limit order book. We also identify conditions under which leverage management practices constitute a sufficient contagion channel for flash crash propagation and \emph{default cascades}.

Our results contribute to the growing literature on the use of ABMs to investigate the complex, emergent dynamics of financial systems \citep{lebaron2000agent, tesfatsion2002agent, farmer2009economy, haldane2011systemic, fagiolo2017,haldane2018interdisciplinary}. ABMs are able to reproduce empirically observed features \citep[\emph{stylised facts},][]{cont2001empirical} of financial systems that cannot easily be captured within mainstream \emph{Dynamic Stochastic General Equilibrium} (DSGE) approaches \citep{fagiolo2017}. The complex interactions of heterogeneous, boundedly-rational, autonomous agents give rise to \emph{emergent phenomena} that cannot be predicted from an understanding of agent behaviour alone. However, this expressiveness comes at a cost. ABMs typically feature a large number of parameters, making model calibration difficult and costly in terms of computational resources. It is, furthermore, necessary to perform repeated stochastic \emph{Monte Carlo} simulations of ABMs in order to build statistical confidence in observed system dynamics \citep{franke2012structural}. However, the financial crisis of 2007-2009 demonstrated the inadequacy of mainstream DSGE models in analysing the systemic risk of the banking system \citep{fagiolo2017}. ABMs offer a compelling alternative, and advances in their methodology and in computer power and data availability will further enhance their effectiveness.

This paper proceeds as follows. Section \ref{sec:related_work} discusses related studies. Section \ref{sec:methods} presents our joint microscopic/macroscopic simulation methodology. Section \ref{sub:empirical_validation_results} presents an empirical validation of the model. Section \ref{sub:effect_of_asset_shock_magnitude} presents results demonstrating the effect of distressed selling on asset prices. Results concerning the impact of leverage, capitalisation and margin tolerance on systemic risk are provided in Section \ref{sub:effect_of_leverage}. The effect of fund--asset network topology on systemic risk is investigated in Section \ref{sub:effect_of_network_topology}. Section \ref{sec:discussion_and_further_work} concludes the paper and presents further work directions.

\section{Related Work}
\label{sec:related_work}

There has been growing recognition of the importance of considering network effects as contributing factors to systemic risk in the financial industry e.g., \citet{gai2010contagion, gai2011complexity, haldane2011systemic,acemoglu2015systemic,haldane2018interdisciplinary}. Within this literature, distressed deleveraging \citep[fire sales,][]{shleifer2011fire, greenwood2015vulnerable}, has been established as a major channel for financial distress contagion via shared asset holdings \citep{arinaminpathy2012size, caccioli2014stability, caccioli2015overlapping,  battiston2016leveraging, capponi2015price, cont2017fire}. When funds hold overlapping portfolios of shared assets this is known as \emph{crowding} \citep{menkveld2015crowded, sias2016hedge}. Price impact \citep{toth2011anomalous, zarinelli2015beyond} is critical to understanding the role of position crowding in fire-sale distress. Price impact under distressed market conditions has been shown to exhibit very different properties to price impact under equilibrium conditions \citep{cristelli2010liquidity}. However, as Table \ref{table:linear_impact} illustrates, this aspect of shared-asset models is often treated in a simplified manner. This point is also raised by \citet{benoit2016risks}. Table \ref{table:linear_impact} presents a range of studies from the macroscopic systemic risk literature, each of which features overlapping asset portfolios as a contagion channel for financial distress. We present the method of price impact calculation deployed in the papers. Although the table features both linear and exponential forms for price impact, the two forms are closely related. As \citet[p.238]{caccioli2014stability} states, exponential price impact ``corresponds to linear market impact for log-prices''. The papers surveyed contain a mix of single- and multi-asset models. Some of these multi-asset models combine assets into representative clusters which are referred to as \emph{asset classes}. Models with a single asset are unable to investigate the propagation of distress between assets, such as that encountered during flash crashes. Models that use a representative asset class are unable to simulate the price dynamics of single assets and so the details of liquidity provision are precluded from investigation. Finally, we note in Table \ref{table:linear_impact} that the majority of papers surveyed do not consider \emph{short selling} (the action of borrowing assets to sell in the hope of repurchasing at a lower future price, equivalent to taking negative positions). Such \emph{long-only} models are therefore precluded from investigating financial distress amplification mechanisms such as the \emph{short squeeze} \citep{dechow2001shortsqueeze}.

\begin{table}[h]
\centering
\begin{tabular}{ l l l l}
\toprule
Paper & Price impact & Securities & Investment \\
\midrule
\citet{nier2007network} & exponential\textdagger & single & long\\
\citet{gai2010contagion} & exponential\textdagger & single & long\\
\citet{arinaminpathy2012size} & exponential\textdagger& multiple\ddag& long\\
\citet{cont2013running} & linear & multiple & long/short\\
\citet{huang2013cascading} & linear & multiple & long\\
\citet{caccioli2014stability} & exponential\textdagger & multiple& long\\
\citet{di2015assessing} & linear & multiple\ddag& long\\
\citet{greenwood2015vulnerable} & linear & multiple & long\\
\citet{caccioli2015overlapping} & linear & single& long\\
\citet{battiston2016leveraging} & linear & multiple & long\\
\citet{serri2016interbank} & linear & multiple & long\\
\citet{cont2017fire} & linear & multiple & long\\
\bottomrule
\end{tabular}
\caption{A survey of price impact functions and portfolio investment constraints utilised in shared-asset models of financial contagion. \textdagger corresponds to ``linear market impact for log-prices'' \citep[p.238]{caccioli2014stability}. \ddag represents the set of equities or bonds (for example) by a single representative \emph{asset class}. \emph{Long} and \emph{short} refer respectively to positively and negatively signed investments.}
\label{table:linear_impact}
\end{table}

Linear price impact functions fail to capture the emergent and time-dependent nature of market liquidity (the ability to freely buy and sell assets at current prices), the study of which falls within the market microstructure literature \citep{gould2013lobs}. Events such as the Flash Crash of 6th May 2010 \citep{kirilenko2017flash} demonstrate that price impact during crisis periods can be extreme. On that occasion, asset prices exhibited unprecedented volatility as the orderly provision of market liquidity broke down. The event was also characterised by the high speed at which prices fell and subsequently returned almost to pre-crash levels. Automated \emph{algorithmic trading}  systems were implicated in all phases of the Flash Crash, including triggering the event, sustaining price falls, and facilitating eventual reversals \citep{cftc2010findings}.

Two key and related mechanisms mean that crisis price impact is different from equilibrium price impact. \citet{cristelli2010liquidity} and \citet{bookstaber2015agent} argue that, during crises, market participants refrain from placing orders. \citet{easley2012flow} relate this phenomenon to the inferred presence of \emph{toxic order flow}, i.e. strongly directional price movements that are outside non-crisis bounds. This removal of liquidity leads to sparsely populated LOBs, and as discussed by \citet{farmer2004really}, gaps on the LOB can lead to further large price movements.

Recent research has demonstrated the success of agent-based approaches in investigating intraday market stability, and in particular the role of HFT \citep{menkveld2016economics} in liquidity provision and flash crashes e.g. \citet{hanson2011effects, johnson2012financial, paddrik2012agent, vuorenmaa2014agent, budish2015high, jacob2017market}. Although these models are successful at reproducing stylized facts of financial markets including during crises, attempts to extend these models to multi-asset or multi-venue settings have been extremely limited.

\citet{kirilenko2017flash} and \citet{golub2012high} take purely empirical approaches to understanding the causes of the Flash Crash of 6th May 2010. \citet{kirilenko2017flash} use regression analysis on a unique dataset labelled with the identities of market participants. This analysis demonstrated that HFT activity resulted in a ``hot-potato'' effect exacerbating price declines. \citet{golub2012high} looks at order placement activity in markets one minute either side of mini-flash crash events \citep{johnson2012financial} finding HFTs withdraw liquidity when it is needed most, at the same time as market makers and other liquidity providers.

\citet{hanson2011effects}, \citet{paddrik2012agent}, \citet{vuorenmaa2014agent}, and \citet{jacob2017market} use combinations of agents representing low and high frequency market participants and are successful at reproducing the characteristic price fall and recovery observed during flash crashes. Indeed, the price recovery observed during flash crashes is another way in which crisis price impact differs from the linear models used in the fire-sales literature. During the Flash Crash of 6th May 2010, financial distress was observed to spread throughout equity and futures markets, making it a systemic event. However, microstructure models typically consider a single asset in isolation and do not incorporate network effects. This is a significant limitation of current approaches, and further research is necessary if the systemic risk due to flash crashes is to be successfully modelled and analysed
\citet{hanson2011effects,paddrik2012agent,vuorenmaa2014agent,jacob2017market}.

Microstructure models featuring multiple assets are rare in the literature. \citet{gerig2015high} takes a model of informed (fundamentalist) and uninformed (zero-intelligence) traders across multiple assets and finds that stocks with higher simulated HFT activity are more highly correlated. The authors also find that close price coupling allows pricing errors to propagate more readily. However, the model does not specifically consider network topology, financial stability, systemic risk, or flash crashes. \citet{torii2015shock} consider a three-asset agent-based model of
arbitrage where an \emph{index future} (a financial product constructed to track the combined price movement of a set of underlying assets) is traded simultaneously with the set of underlying assets. \citet{torii2015shock} find that price shocks can propagate from asset to asset via movements in the index future. The model does not consider the topology of the asset-index network, or flash crashes. However, this is a notable result given the paucity of microstructure models considering the contagion of financial distress. \citet{cespa2014illiquidity} consider a two-asset model for distress propagation where an ``illiquidity cycle'' emerges as traders avoid changing their investments when price volatility leads to uncertainty and a reduced ``risk appetite''. As with other closed-form approaches, this model does not feature a realistic model of order placement activity.

The purpose of the present paper is to provide a realistic microstructure model of price impact during crisis periods, embedded within a multiple fund, multiple asset environment characteristic of models from the macroscopic systemic risk literature. The model allows us to explore the propagation of financial distress associated with inherently intraday market phenomena such as flash crashes. Events taking place over timescales measured in minutes, seconds, or shorter are simply ignored in the overwhelming majority of macroscopic financial models. However, the impact of such events can, as illustrated by the 6th May 2010 Flash Crash, be measured in trillions of dollars and serve to undermine the orderly functioning of global financial markets.

\section{A Micro-Macro Model}
\label{sec:methods}

Our model features the novel synthesis of a macroscopic fund-asset network with a microscopic limit order book trading simulation. When funds adjust their asset holdings, they do so by placing limit orders in a realistically calibrated stock market model, where heterogeneous trading agents interact with the fund's orders. Price impact arises endogenously from our simulation, and resulting asset price movements affect the profits and losses of the funds.
In order to represent the composite micro-macro structure it is necessary to introduce a set of heterogeneous interacting entities:

\begin{enumerate}
  \item Banks: to model the provision of leveraged financing to institutions
  \item Funds: to model institutions holding leveraged positions in risky assets
  \item Assets: to model liquidity provision via the limit order book
  \item Traders: to model realistic asset market activity and response to market distress
\end{enumerate}

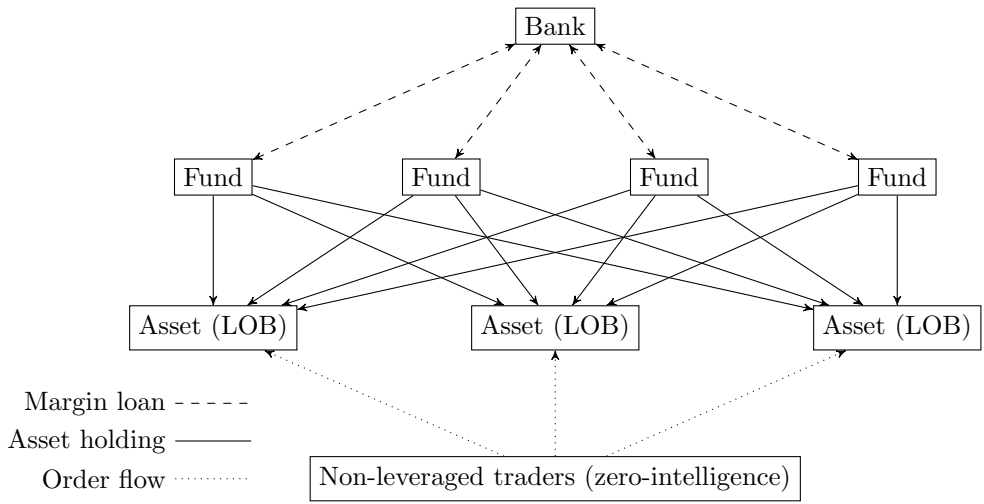
\begin{figure}
\centering
\begin{tikzpicture}

  \node [draw, shape=rectangle] (bank) at (5,4) {Bank};

  \node [draw, shape=rectangle] (fund0) at (0.5,2) {Fund};
  \node [draw, shape=rectangle] (fund1) at (3.5,2) {Fund};
  \node [draw, shape=rectangle] (fund2) at (6.5,2) {Fund};
  \node [draw, shape=rectangle] (fund3) at (9.5,2) {Fund};

  \node [draw, shape=rectangle] (asset0) at (0.5,0) {Asset (LOB)};
  \node [draw, shape=rectangle] (asset1) at (5,0) {Asset (LOB)};
  \node [draw, shape=rectangle] (asset2) at (9.5,0) {Asset (LOB)};

  \node [draw, shape=rectangle] (traders) at (5,-2) {Non-leveraged traders (zero-intelligence)};

  \draw [<->, dashed] (bank) -- (fund0);
  \draw [<->, dashed] (bank) -- (fund1);
  \draw [<->, dashed] (bank) -- (fund2);
  \draw [<->, dashed] (bank) -- (fund3);

  \draw [->] (fund0) -- (asset0);
  \draw [->] (fund0) -- (asset1);
  \draw [->] (fund0) -- (asset2);

  \draw [->] (fund1) -- (asset0);
  \draw [->] (fund1) -- (asset1);
  \draw [->] (fund1) -- (asset2);

  \draw [->] (fund2) -- (asset0);
  \draw [->] (fund2) -- (asset1);
  \draw [->] (fund2) -- (asset2);

  \draw [->] (fund3) -- (asset0);
  \draw [->] (fund3) -- (asset1);
  \draw [->] (fund3) -- (asset2);

  \draw [->, dotted] (traders) -- (asset0);
  \draw [->, dotted] (traders) -- (asset1);
  \draw [->, dotted] (traders) -- (asset2);

  \draw [dashed] (0, -1) -- (1, -1) node [pos=0.0, left] {Margin loan};
  \draw [] (0, -1.5) -- (1, -1.5) node [pos=0.0, left] {Asset holding};
  \draw [dotted] (0, -2) -- (1, -2) node [pos=0.0, left] {Order flow};

\end{tikzpicture}
\caption{\singlehalf Relationships between entities in the model. Arrows represent balance-sheet
relationships. Between banks and funds, these relationships refer to secured
margin loans and leverage. Between funds and assets, the relationships refer
to asset ownership within fund portfolios. Similarly, between traders and assets
the relationship again refers to asset ownership in trader portfolios, though
in the present model we assume that traders are non-leveraged and so not subject
to margin calls. Note that we do not model banks directly investing in assets.}
\label{fig:interactions}
\end{figure}

A summary of the agents and other entities in the model is presented in Figure
\ref{fig:interactions}. Dashed arrows indicate secured margin loans extended
by the single bank to the fund agents (for the sake of clarity only a small number
of funds and assets are represented). Solid arrows represent fund holdings in
assets, and this section of the system is what we refer to throughout this work
as the ``bipartite fund-asset network''. Finally, dotted arrows indicate the
order placement activity of the non-leveraged trader agents described in the
previous section.

In the sections that follow we will discuss how we capture these entities in
our model, but we first present the key assumptions underpinning the present work.

\subsection{Assumptions}
\label{sub:assumptions}

Before we describe in detail the components of our model, it is first necessary to establish
the terms of what the model does and does not set out to simulate.

\subsubsection{Asset-related assumptions}
\label{ssub:asset_related_assumptions}

Asset microstructure is represented using limit order books. Transactions occur when orders to buy (sell) match against previously submitted orders to sell (buy) \citep{gould2013lobs}. Our asset model is supported by the following key assumptions.

We assume:

\begin{itemize}
    \item a single currency for all financial transactions in the model.
    \item that trading is costless in the sense that there
are no commissions, exchange fees or rebates involved in the model. Such costs
are likely to be immaterial in comparison to the extreme price impact encountered
during flash crash type phenomena.
    \item that assets have no corporate actions such as splits or dividends (a minor
assumption given the timescale of corporate actions is weeks rather than intraday).
    \item that assets have the
same liquidity and trading characteristics, as well as holding the same start
price at the beginning of simulation runs. The liquidity assumption can be compared to restricting our asset universe to members of a major index such as the Eurostoxx 50 or S\&P 500.
    \item that assets are not directly linked via derivative contracts such as options or via direct asset--asset ownership. Such links introduce new potential channels for distress propagation and so our model is conservative in this regard.
    \item that all trading takes place at a single trading venue. Market participant confusion over conflicting data from alternative trading venues was implicated in the Flash Crash of 6th May 2010, and so again our model represents a conservative configuration \citep{cftc2010findings}.
\end{itemize}

\subsubsection{Fund-related assumptions}
\label{ssub:fund_related_assumptions}

Our research questions relate to how the fund-asset network impacts systemic
risk (sections \ref{sub:effect_of_leverage} and \ref{sub:effect_of_network_topology}), in the present work we therefore do not concern ourselves with how funds build positions
first. We characterise the funds in the model as large institutions
that take a long time (many days at least) to accumulate positions, and so over
the short intraday timescales we wish to consider, such as the duration of a flash
crash (minutes), their
pre-crisis behaviour is irrelevant to our research questions. However, if a fund
enters distress via the enactment of a margin call, it becomes active and seeks
to rapidly reduce leverage.

We assume:

\begin{itemize}
    \item that funds already hold a set of asset positions before the simulation
begins and do not adjust positions unless experiencing financial distress.
    \item that funds do not invest directly in other funds (so-called \emph{fund-of-funds}). As with asset-asset direct linkage, our model takes a conservative stance in this regard.
    \item that funds deploy all of their capital as collateral
for leverage loans and so have no cash reserves available in order to satisfy margin
calls. This implies that funds must liquidate (reduce in absolute terms) asset positions in order to
reduce leverage. Capital injection is equivalent to increased margin tolerance, which we investigate comprehensively in the present paper.
    \item that funds liquidate aggressively (quickly, by taking liquidity), similarly to
the empirical observations by \citet{adrian2010liquidity} and the model
by \citet{greenwood2015vulnerable}.
    \item that when funds liquidate, they liquidate uniformly across their assets. This approach follows the methodology established by \citet{caccioli2014stability}.
    \item that all funds in
the simulation have equal starting capital \citep[also as in][]{caccioli2014stability}.
    \item that, in line with established methodology (see Table \ref{table:linear_impact}),
asset positions are long-only (positive). This strong assumption means that
our present work fits into the fire sales literature \citep{shleifer2011fire, greenwood2015vulnerable}
and does not attempt to explain long-short phenomena such as the 2007 ``Quant Meltdown''
\citep{khandani2007happened}\footnote{\citet{khandani2007happened} and \citet{cont2013running} argue that crowding at the trading strategy \citep[or \emph{risk factor},][]{fama93commonrisk} level describes the financial contagion observed during the ``Quant Meltdown'' more clearly than consideration of individual investments in single financial securities.}.
\end{itemize}

\subsubsection{Bank-related assumptions}
\label{ssub:bank_related_assumptions}

In our model, financial distress
propagates between funds connected via shared asset holdings. It is not, therefore,
necessary to consider interbank contagion in order to address our research questions.
Neither is it necessary for us to represent banks as directly owning assets; this
behaviour is encapsulated entirely within funds and so provides a clear separation
of responsibilities without loss of generality.
Our model thus incorporates a single bank that provides leverage loans to all funds in
the simulation.

We assume:

\begin{itemize}
    \item that all funds in the simulation are offered equal leverage terms, following established methodology \citep{caccioli2014stability}.
    \item that fund portfolio constitution is ignored when banks set leverage
limits. In reality, different assets attract different levels of permissible
leverage. Increased portfolio volatility reduces the cap and so represents an additional channel of distress propagation. Our model is conservative with respect to this effect, in line with the implicit methodology of related models.
    \item that banks monitor fund portfolios on an intraday basis and issue margin
calls to funds whose current leverage breaches some threshold. This assumption is realistic \citep{brunnermeier2009market,cont2017fire}.
    \item that if a fund defaults (trading losses exceed capital) the bank will take over and liquidate the fund's portfolio. Margin
terms typically provide for this repossession.
    \item that there are no other
funding costs or interest charged by banks to funds.
We justify this on the basis of timescale --- such fees are, in reality, accounted on a daily
or monthly basis and are not material to the short intraday timescales relevant
to our research questions.
\end{itemize}

\subsection{Agent-Based Model}
\label{sub:structure_of_the_model}

In order to simulate the emergent properties of the interconnected, heterogeneous
system of banks, funds, assets and traders we implement an agent-based model (ABM). The ABM
consists of software agents representing banks, funds and traders. The model further incorporates
assets and the markets for the trading of those assets but these are not considered
agents in their own right. Our notion of which entities in the model constitute
agents follows \citet[p.21]{wooldridge2001multi}:

\begin{displayquote}
``An \emph{agent} is a computer system that is \emph{situated} in some \emph{environment}, and that
is capable of \emph{autonomous action} in this environment in order to meet its delegated objectives.''
\end{displayquote}

In the sections that follow we describe the behaviour of, and interconnections
between, agents in the model.

\subsubsection{Banks}
\label{ssub:banks}

As discussed above, we model the activity of a single bank whose primary purpose
is to provide leveraged financing to funds. The bank is modelled by considering
its balance sheet of assets and liabilities, in line with existing approaches
\citep{gai2010contagion}. The balance sheet for the bank in our model is presented
in Table \ref{table:balance_sheet_bank}.

\begin{table}[h]
\centering
\begin{tabular}{ c  c }
\toprule
Assets & Liabilities \\
\midrule
Margin lending     & Loss on margin loans \\
Cash               & Equity \\
\bottomrule
\end{tabular}
\caption{Bank balance sheet model. Equity is defined as the sum of assets minus
the sum of liabilities.}
\label{table:balance_sheet_bank}
\end{table}

As shown in Table \ref{table:balance_sheet_bank} and discussed previously, we
do not model interbank lending in the present work. The function of the bank in
our model therefore is to issue debt to funds
which is secured using the fund's initial capital $C^0 \in \mathbb{R}^+$ as collateral. Banks
offer funds identical initial leverage $\lambda^0_i = \lambda^0, \forall i \in \{1, ..., n_f\}$,
$\lambda^0 \in \mathbb{R}^+$.
The total value $L_i$ of a loan made by the bank to the $i$th fund is
therefore given by $L_i = C^0_i\lambda^0_i$. This expression demonstrates an important convention regarding leverage: the total cash
available to funds for the purposes of investment, $V$, is given by $V_i = L_i + C^0_i = C^0_i(\lambda^0_i + 1)$, so leverage of zero ($\lambda^0=0$) corresponds to the case where $L_i=0$, that
is, no cash is loaned.

As explained above, we assume that each of the $n_f$ funds utilise all of their available capital
for the purposes of funding asset positions. If we represent the investment
of fund $i$ in asset $j$ as elements of a matrix $A_{ij}$ then we may write
$C^0_i(\lambda^0_i + 1) = \sum_{j=1}^{n_a}A_{ij}$. If we further represent asset positions as the product of asset price at time $t$, $p_j^t \in \mathbb{R}^+$, and
position size in shares, $S_{ij}^t \in \mathbb{R}$, then for
time $t=0$ we can write $C^0_i(\lambda^0_i + 1) = \sum_{j=1}^{n_a}S_{ij}^0p_j^0$, which makes explicit the relationship between leverage, capital, price and fund asset holdings.
As prices fluctuate the total fund portfolio value will also of course vary. This
price variation will result in mark-to-market profit and loss at various times accruing to the fund.
This profit and loss is combined with the fund's capital yielding an updated
capital at time $t$, $C_i^t$. We can thus derive the
leverage for fund $i$ at time $t$, $\lambda^t_i$, which is given by

\[
\lambda^t_i = \frac{1}{C^t_i} \sum_{j=1}^{n_a}S_{ij}^tp_j^t - 1.
\]

As in \citet{cont2017fire}, we assume that banks continuously measure the ratio $\tau = \sfrac{\lambda^t}{\lambda^0}$
for all funds and issue margin calls when this ratio exceeds a critical hysteresis
threshold $\tau_c \in \lbrack1, \infty)$ (where the infinite limit implies
banks never issue margin calls, and the lower limit of 1 implies banks issue
margin calls for infinitesimally small fund losses). The margin call state
remains in force at a fund until either the fund reduces its leverage $\lambda_i^t$
such that $\lambda_i^t < \lambda_i^0$, or the fund defaults (which occurs when $C^t_i < 0$). Note that in our model the effect of a bank issuing a margin call to a fund is exactly equivalent to a fund managing its leverage to a target -- a common industry practice \citep{cont2017fire}.
If a fund defaults,
we invoke the bank's realistic prerogative to seize its assets and to continue to liquidate the positions.

Banks act autonomously in their decision as to whether and when to issue margin calls. Although we do not
explore bank decision-making in the present work, the possibility justifies our consideration
of banks as agents proper in the model.

\subsubsection{Funds}
\label{ssub:funds}

The primary purpose of funds in the model is to accept leverage loans from banks and to
use these loans to finance a portfolio of assets (see Section \ref{ssub:banks}).
The model incorporates many funds, each with its own randomly allocated portfolio
(details of the allocation process are provided below). As asset prices vary
as a result of the microstructure simulation described in the next section, so
the leverage of funds also varies as profit and loss accrues to the fund's
account. If leverage increases beyond a critical threshold (see Section \ref{ssub:banks}
for the mathematical details) the fund will enter a margin call state and will
aggressively liquidate its portfolio (details of the liquidation mechanism and how this
relates to the microstructure model are presented in Section \ref{ssub:traders}
below). Funds can, in principle, decide autonomously which assets
they elect to liquidate. Indeed, \citet{capponi2015price} find that the strategic
selection of asset liquidation behaviour has a large impact on systemic risk in
a macroscopic model. We do not consider this strategic behaviour in the present work. Further fund autonomy manifests in the selection of distressed liquidation
parameters, and we explore systemic sensitivity to this in Section \ref{sub:effect_of_asset_shock_magnitude}.

Funds are represented by their balance sheet, as shown in Table \ref{table:balance_sheet_fund}.
As described previously, we do not simulate the acquisition of fund asset
portfolios, rather we specify portfolios as part of the initial conditions of
simulation runs. This approach allows us to explore portfolios with specific
characteristics and their effect on systemic risk.

\begin{table}[h]
\centering
\begin{tabular}{c  c}
\toprule
Assets & Liabilities \\
\midrule
Capital & Margin loan \\
Portfolio of assets & Portfolio profit/loss \\
& Equity \\
\bottomrule
\end{tabular}
\caption{Fund balance sheet model}
\label{table:balance_sheet_fund}
\end{table}

\subsubsection{Fund-Asset Bipartite Network}
\label{ssub:fund_asset_bipartite_network}

Following the approaches of \citet{caccioli2014stability} and \citet{huang2013cascading} we define the random bipartite graph \citep{newman2003structure} $\mathcal{G}=(X \cup Y, E)$ in which edges $E$ represent investments made by funds (bottom nodes, $X$) in assets (top nodes, $Y$) such that $|X|=n_f$ and $|Y|=n_a$. The structure of $\mathcal{G}$ can be represented by the matrix $A \in \mathbb{R}^{n_f\times n_a}$, defined in Section \ref{ssub:banks}. A non-zero element $A_{ij}$ correspond to the presence of an edge $E \in \mathcal{G}$ between fund $i$ and asset $j$. $A$ is thus the \emph{adjacency matrix} representation of $\mathcal{G}$ \citep{newman2003structure}.

We introduce two parameters to control the topology of $\mathcal{G}$. The fraction of assets in which a fund invests \citep[the \emph{diversification}, or \emph{density},][]{caccioli2014stability} is controlled by parameter $\rho \in [0,1], \rho \in \mathbb{R}$. Funds
are restricted such that they hold investments in at least one asset, that is,
un-weighted fund degree $k_{\mathit{fund}} = \text{max}(\lfloor \rho n_a \rfloor, 1)$. Since the same value of $\rho$ is used for all funds,
as $\rho \to 1$, all funds converge to the same
portfolio (such that every fund invests in every asset). Conversely, $\rho \to 0$ means each fund invests in a single asset.

In order to generate networks
with a controllable level of overlap, we introduce the notion
of preferential attachment \citep{barabasi1999preferential} governed by a parameter $\beta \in \mathbb{R}$ controlling the non-linear strength of preferential attachment \citep{krapivsky2000connectivity}.
Starting with the set of assets, network construction proceeds by adding funds sequentially. Each additional fund makes weighted connections to $k_{\mathit{fund}}$ different assets. The probability that the fund connects with a given asset is related to the total already invested in that asset, modulated by crowding parameter $\beta$. $\beta = 0$ generates a fully stochastic \erdos random graph \citep{newman2003structure}. $\beta = 1$ generates the linear \ba model, and $\beta > 1$ generates a network in which funds have a very
high probability of selecting assets that are already held by other funds. This can
leave some assets disconnected from the network in our model, and we account for
this in the results that follow. Our model also supports $\beta < 0$,
which generates networks in which funds prefer to invest in assets
with the smallest existing investment. We refer to such network configurations as \emph{dispersed}. Where $n_f \leqslant n_a$ it is possible for each
fund to invest in a different asset producing a maximally disjoint network with
no possibility of contagion between assets. \citet{chen2014asset} find that
such a network is the most stable configuration in high-leverage scenarios and we seek to test
this assertion in our micro-macro model.

Unlike \citet{caccioli2014stability}, we do not assume that funds make equal investments to each asset in their portfolio. Instead, for each fund we sample $k_{\mathit{fund}}$ random variates from a Gaussian distribution $\mathcal{N}(0, 1)$, normalising such that the absolute values of the variates sum to unity. Edges are added between the fund and $k_{\mathit{fund}}$ assets with edge weights assigned in decreasing size order from the set of normalised variates. Algorithm \ref{algo:random_bipartite} in \ref{sec:bipartite} presents the network construction method detail.

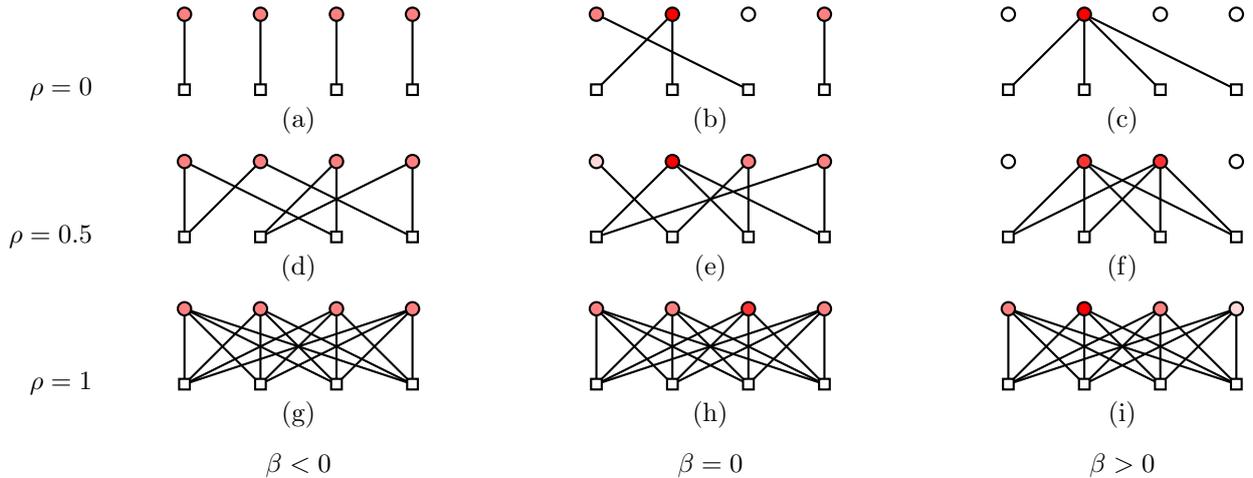
\begin{figure}
\centering

  \begin{tabular*}{\textwidth}{@{\extracolsep{\fill}}rccccc}

    $\rho=0$ &

    \begin{tikzpicture}
    \draw [thick] (0, 0) -- (0, 1);
    \draw [thick] (1, 0) -- (1, 1);
    \draw [thick] (2, 0) -- (2, 1);
    \draw [thick] (3, 0) -- (3, 1);
    \draw [thick, black,fill=white] plot [only marks, mark=square*] coordinates {(0,0) (1,0) (2,0) (3,0)};
    \draw [thick, black,fill=red!50!white] plot [only marks, mark size=2.5, mark=*] coordinates {(0,1) (1,1) (2,1) (3,1)};
    \end{tikzpicture}

    & &

    \begin{tikzpicture}
    \draw [thick] (0, 0) -- (1, 1);
    \draw [thick] (1, 0) -- (1, 1);
    \draw [thick] (2, 0) -- (0, 1);
    \draw [thick] (3, 0) -- (3, 1);
    \draw [thick, black,fill=white] plot [only marks, mark=square*] coordinates {(0,0) (1,0) (2,0) (3,0)};

    \draw [thick, black,fill=white] plot [only marks, mark size=2.5, mark=*] coordinates {(2,1)};
    \draw [thick, black,fill=red!50!white] plot [only marks, mark size=2.5, mark=*] coordinates {(0,1) (3,1)};
    \draw [thick, black,fill=red] plot [only marks, mark size=2.5, mark=*] coordinates {(1,1)};
    \end{tikzpicture}
    & &

    \begin{tikzpicture}
    \draw [thick] (0, 0) -- (1, 1);
    \draw [thick] (1, 0) -- (1, 1);
    \draw [thick] (2, 0) -- (1, 1);
    \draw [thick] (3, 0) -- (1, 1);
    \draw [thick, black,fill=white] plot [only marks, mark=square*] coordinates {(0,0) (1,0) (2,0) (3,0)};
    \draw [thick, black,fill=white] plot [only marks, mark size=2.5, mark=*] coordinates {(0,1) (2,1) (3,1)};
    \draw [thick, black,fill=red] plot [only marks, mark size=2.5, mark=*] coordinates {(1,1)};
    \end{tikzpicture}

    \\

     & (a) & &  (b) &  & (c) \\ [7pt]

    $\rho=0.5$ &

    \begin{tikzpicture}
    \draw [thick] (0, 0) -- (0, 1);
    \draw [thick] (0, 0) -- (1, 1);

    \draw [thick] (1, 0) -- (2, 1);
    \draw [thick] (1, 0) -- (3, 1);

    \draw [thick] (2, 0) -- (0, 1);
    \draw [thick] (2, 0) -- (2, 1);

    \draw [thick] (3, 0) -- (1, 1);
    \draw [thick] (3, 0) -- (3, 1);
    \draw [thick, black,fill=white] plot [only marks, mark=square*] coordinates {(0,0) (1,0) (2,0) (3,0)};
    \draw [thick, black,fill=red!50!white] plot [only marks, mark size=2.5, mark=*] coordinates {(0,1) (1,1) (2,1) (3,1)};
    \end{tikzpicture}

    & &

    \begin{tikzpicture}
    \draw [thick] (0, 0) -- (1, 1);
    \draw [thick] (0, 0) -- (3, 1);

    \draw [thick] (1, 0) -- (0, 1);
    \draw [thick] (1, 0) -- (2, 1);

    \draw [thick] (2, 0) -- (1, 1);
    \draw [thick] (2, 0) -- (2, 1);

    \draw [thick] (3, 0) -- (1, 1);
    \draw [thick] (3, 0) -- (3, 1);

    \draw [thick, black,fill=white] plot [only marks, mark=square*] coordinates {(0,0) (1,0) (2,0) (3,0)};
    \draw [thick, black,fill=red] plot [only marks, mark size=2.5, mark=*] coordinates {(1,1)};
    \draw [thick, black,fill=red!50!white] plot [only marks, mark size=2.5, mark=*] coordinates {(2,1) (3,1)};
    \draw [thick, black,fill=red!15!white] plot [only marks, mark size=2.5, mark=*] coordinates {(0,1)};
    \end{tikzpicture}

    & &

    \begin{tikzpicture}
    \draw [thick] (0, 0) -- (1, 1);
    \draw [thick] (0, 0) -- (2, 1);

    \draw [thick] (1, 0) -- (1, 1);
    \draw [thick] (1, 0) -- (2, 1);

    \draw [thick] (2, 0) -- (1, 1);
    \draw [thick] (2, 0) -- (2, 1);

    \draw [thick] (3, 0) -- (1, 1);
    \draw [thick] (3, 0) -- (2, 1);

    \draw [thick, black,fill=white] plot [only marks, mark=square*] coordinates {(0,0) (1,0) (2,0) (3,0)};
    \draw [thick, black,fill=white] plot [only marks, mark size=2.5, mark=*] coordinates {(0,1)  (3,1)};
    \draw [thick, black,fill=red!80!white] plot [only marks, mark size=2.5, mark=*] coordinates {(1,1) (2,1)};
    \end{tikzpicture}

    \\

    & (d)  & & (e) &  & (f) \\ [7pt]

    $\rho=1$ &

    \begin{tikzpicture}
    \draw [thick] (0, 0) -- (0, 1);
    \draw [thick] (0, 0) -- (1, 1);
    \draw [thick] (0, 0) -- (2, 1);
    \draw [thick] (0, 0) -- (3, 1);

    \draw [thick] (1, 0) -- (0, 1);
    \draw [thick] (1, 0) -- (1, 1);
    \draw [thick] (1, 0) -- (2, 1);
    \draw [thick] (1, 0) -- (3, 1);

    \draw [thick] (2, 0) -- (0, 1);
    \draw [thick] (2, 0) -- (1, 1);
    \draw [thick] (2, 0) -- (2, 1);
    \draw [thick] (2, 0) -- (3, 1);

    \draw [thick] (3, 0) -- (0, 1);
    \draw [thick] (3, 0) -- (1, 1);
    \draw [thick] (3, 0) -- (2, 1);
    \draw [thick] (3, 0) -- (3, 1);

    \draw [thick, black,fill=white] plot [only marks, mark=square*] coordinates {(0,0) (1,0) (2,0) (3,0)};
    \draw [thick, black,fill=red!50!white] plot [only marks, mark size=2.5, mark=*] coordinates {(0,1) (1,1) (2,1) (3,1)};
    \end{tikzpicture}

    & &

    \begin{tikzpicture}
    \draw [thick] (0, 0) -- (0, 1);
    \draw [thick] (0, 0) -- (1, 1);
    \draw [thick] (0, 0) -- (2, 1);
    \draw [thick] (0, 0) -- (3, 1);

    \draw [thick] (1, 0) -- (0, 1);
    \draw [thick] (1, 0) -- (1, 1);
    \draw [thick] (1, 0) -- (2, 1);
    \draw [thick] (1, 0) -- (3, 1);

    \draw [thick] (2, 0) -- (0, 1);
    \draw [thick] (2, 0) -- (1, 1);
    \draw [thick] (2, 0) -- (2, 1);
    \draw [thick] (2, 0) -- (3, 1);

    \draw [thick] (3, 0) -- (0, 1);
    \draw [thick] (3, 0) -- (1, 1);
    \draw [thick] (3, 0) -- (2, 1);
    \draw [thick] (3, 0) -- (3, 1);
    \draw [thick, black,fill=white] plot [only marks, mark=square*] coordinates {(0,0) (1,0) (2,0) (3,0)};
    \draw [thick, black,fill=red!50!white] plot [only marks, mark size=2.5, mark=*] coordinates {(0,1) (1,1) (3,1)};
    \draw [thick, black,fill=red!80!white] plot [only marks, mark size=2.5, mark=*] coordinates {(2,1)};
    \end{tikzpicture}

    & &

    \begin{tikzpicture}
    \draw [thick] (0, 0) -- (0, 1);
    \draw [thick] (0, 0) -- (1, 1);
    \draw [thick] (0, 0) -- (2, 1);
    \draw [thick] (0, 0) -- (3, 1);

    \draw [thick] (1, 0) -- (0, 1);
    \draw [thick] (1, 0) -- (1, 1);
    \draw [thick] (1, 0) -- (2, 1);
    \draw [thick] (1, 0) -- (3, 1);

    \draw [thick] (2, 0) -- (0, 1);
    \draw [thick] (2, 0) -- (1, 1);
    \draw [thick] (2, 0) -- (2, 1);
    \draw [thick] (2, 0) -- (3, 1);

    \draw [thick] (3, 0) -- (0, 1);
    \draw [thick] (3, 0) -- (1, 1);
    \draw [thick] (3, 0) -- (2, 1);
    \draw [thick] (3, 0) -- (3, 1);
    \draw [thick, black,fill=white] plot [only marks, mark=square*] coordinates {(0,0) (1,0) (2,0) (3,0)};
    \draw [thick, black,fill=red!50!white] plot [only marks, mark size=2.5, mark=*] coordinates {(0,1)  (2,1) };
    \draw [thick, black,fill=red] plot [only marks, mark size=2.5, mark=*] coordinates {(1,1)};
    \draw [thick, black,fill=red!15!white] plot [only marks, mark size=2.5, mark=*] coordinates {(3,1)};
    \end{tikzpicture}

    \\

     & (g) &  & (h) & &  (i) \\ [7pt]

    & $\beta < 0$ &  & $\beta = 0$ &  & $\beta > 0$ \\ [7pt]

  \end{tabular*}

\caption{\double \incolour Illustrative examples of the range of bipartite topologies that can be generated with our network construction method. Top nodes (circles) represent assets, bottom nodes (squares) represent funds. An edge between two nodes represents asset ownership by a fund. Panels on the same row have equal diversification parameter $\rho$, and panels in the same column have equal crowding parameter $\beta$. Darker asset node shading indicates a higher proportional amount invested in the asset.}
\label{fig:bipartite_diagrams}

\end{figure}

Figure \ref{fig:bipartite_diagrams} demonstrates the range of topologies that can be created using our bipartite preferential attachment method (for clarity we present small networks in which $n_f = n_a = 4$). Panels on the same row have equal diversification parameter $\rho$, and panels in the same column have equal crowding parameter $\beta$. Figure \ref{fig:bipartite_diagrams}(a) shows a low density, dispersed configuration in which each fund holds a different asset. This network is disconnected, prohibiting default cascades. As crowding parameter $\beta$ increases in Figure \ref{fig:bipartite_diagrams}(b), the network becomes somewhat more connected with two funds sharing a common asset. The effect of preferential attachment becomes apparent in Figure \ref{fig:bipartite_diagrams}(c) where all funds invest in the same single asset, leaving most assets disconnected from the network. Figures \ref{fig:bipartite_diagrams}(d--f) illustrate higher diversification networks with $\rho=0.5$. In each case the network consists of a single connected component (disconnected assets are not considered to be network components as they are effectively irrelevant to our simulation). It is possible in connected networks for distress at any fund to affect (directly, or indirectly) all other funds. Finally, Figures \ref{fig:bipartite_diagrams}(g--i) demonstrate that as $\rho \to 1$ networks become \emph{complete} \citep{newman2003structure}. For complete networks it may appear that crowding parameter $\beta$ becomes degenerate, however, for weighted networks this is not necessarily the case. Figures \ref{fig:bipartite_diagrams}(h) and \ref{fig:bipartite_diagrams}(i) demonstrate cases where some nodes receive higher investment than others, even though all nodes have the same (unweighted) degree.

\subsubsection{Traders}
\label{ssub:traders}

Trader agents in our model operate in a single market, single asset capacity in
line with the large majority of the ABM micromarket literature. That is, a single
trader agent is active in a single asset even though it may be controlled by
an entity with interests in multiple assets. Such considerations are of peripheral
importance in the present model since funds do not take an active trading role.
Trader agents, furthermore, are not considered to be leveraged and so will not
be subject to margin calls.

Following well-established principles in the agent-based computational finance
literature, we minimise behavioural assumptions by endowing trader agents with
minimal intelligence. This approach is known as zero-intelligence (ZI), and
is characterised by the lack of forecast capability or adaptation in agents
\citep{gode1993allocative, farmer2005predictive}. Despite their apparent simplicity,
such models are remarkably successful at reproducing stylised facts of financial markets
\citep{fagiolo2007critical, panayi2012agent}.
To be more specific, we reproduce
the model presented in \citet{paddrik2012agent} which features a heterogeneous
set of trader agents, each of which is in turn based upon behavioural observations
of real world trading behaviour on the day of the 6th May 2010 Flash Crash
presented by \citet{kirilenko2017flash}.

Other models were also considered as a
basis for the microscopic LOB simulation \citep{vuorenmaa2014agent, hanson2011effects, jacob2017market}.
All of these models feature a combination of low and high-frequency trading
and so are broadly similar, but the model presented in \citet{paddrik2012agent} had the advantage of removing
any strategic choice on the part of agents and by doing so remaining closer to
the zero-intelligence principle. \citet{paddrik2012agent} has the further advantage
of featuring a realistic order matching mechanism which is closer to real-world
stock exchanges than any of the other models mentioned above.

As in \citet{paddrik2012agent}, we assume agents arrive at the market
according to a Poisson process with a characteristic mean interaction interval
specified per agent type (see Table \ref{table:trader_agents}). This approach
is standard for micromarket models \citep{farmer2005predictive}.

\begin{table}[h]
\centering
\begin{tabular}{l l l l l}
\toprule
Type            & Timescale                                   & Inventory (shares)        & Population                       & Inputs \\
\midrule
Small           & 7200s                                       & $(-\infty, \infty)$       & 6500                             & -- \\
Fundamental B/S & 60s                                         & $(-\infty, \infty)$       & 2500                             & toxic order flow \\
Opportunistic   & 120s                                        & $\lbrack-120, 120\rbrack$ & 1600                             & stochastic signal \\
Market Maker    & 20s                                         & $\lbrack-120, 120\rbrack$ & 320                              & toxic order flow \\
HFT             & 0.35s                                       & $\lbrack-120, 120\rbrack$ & 16                               & LOB imbalance \\
Distressed      & 10s                                         & $(-\infty, \infty)$       & $[1, n_f]$                       & market volume \\
\bottomrule

\end{tabular}
\caption{Trader agents in our model are based on \citet{paddrik2012agent}'s model,
which was calibrated to real market data presented by \citet{kirilenko2017flash}.}
\label{table:trader_agents}
\end{table}

We summarise the main features of the trader agents in Table \ref{table:trader_agents}. Detailed descriptions of the trader agent types are provided in \ref{sec:trader_agent_details} and are closely related to the implementations described in \citep{paddrik2012agent}. However, the present model enhances the role of distressed selling agents from \citet{paddrik2012agent}. The connection between the behaviour of the distressed selling agent and the liquidation behaviour of a controlling fund agent is discussed in the next section.

\subsubsection{Distressed Sellers}

A single aggressive seller executing a large set of trades was identified
as a key causative factor in the 6th May 2010 Flash Crash \citep{kirilenko2017flash}.
Distressed seller agents place market orders (orders with quantity but no price limits)
with an order size limited to a fixed percentage of the previous one-minute trading
volume. This facilitates a positive feedback loop if the placement of
such orders causes market volume to increase, leading to the placement of even
larger orders as observed in the 6th May 2010 Flash Crash.
The distressed sellers in our model come from two sources:

\begin{enumerate}
    \item Exogenous distressed sellers, introduced to shock the system, and
    \item Distressed sellers implementing the liquidation order placement of distressed funds.
\end{enumerate}

First, as in
\citep{paddrik2012agent}, we introduce a distressed seller with a given position
size at a specific time point in the simulation purely to generate an asset price
shock. We allow a period of 20\% of the simulation duration to elapse
before shocking assets to allow initial transient effects to decay. This distressed seller will continue to attempt to liquidate
until it has sold its entire position, after which point it becomes inactive.
The second source of distressed sellers in our model are distressed funds. When
a fund is in a margin call or default state it is compelled to reduce leverage
by liquidating asset positions. We model this by introducing a distressed seller agent
to the market for every asset that the fund chooses to liquidate (in the present
work, this is all assets held by the fund). The distressed seller attempts to
reduce its parent fund's position in each asset to zero. Critically, however, if fund
leverage falls back below the tolerance threshold, liquidation is suspended, and the
distressed agents representing the fund are deactivated. It is possible
that the fund may re-enter distress later and here the distressed selling agents
would reactivate. Our justification for this deleverage mechanism is that margin
call terms usually stipulate that leverage must be brought under control within
a single trading day, thus necessitating immediate and robust action \citep{brunnermeier2009market}.

\subsubsection{Assets}
\label{ssub:assets}

Assets are not treated as agents in the model. Rather, assets represent the
environment in which the other agents are situated \citep{wooldridge2001multi}.
Assets can be bought and sold by trader agents utilising market and limit
orders via a realistic continuous double auction matching engine. The market
clears according to the industry-standard price-time priority \citep{gould2013lobs},
and agents are able to view the current state of the market as represented by
the LOB. Matched orders results in a trade, and the current price of the asset is thus deemed to have been
updated. The simulation is enhanced by the realistic inclusion of an opening
auction period and intraday volatility auctions. In both of these situations,
market clearing is suspended for a period of time to allow liquidity to accumulate
on the LOB and hence aid price discovery. We assume an initial one-minute opening
auction phase and trigger realistically calibrated five-second volatility auctions when
prices fall by 1.3\% in a single second, after \citet{paddrik2012agent}.

\subsection{Systemic Risk Metrics}
\label{sub:systemic_risk_metrics}

For a comprehensive survey of systemic risk metrics the reader is referred to
\citet{bisias2012survey} and \citet{acharya2017measuring}. For the present work
we adopt two notions of systemic risk. The first is tailored to distress at funds, and the
second is tailored to distress in assets.

\subsubsection{Financial Contagion}
\label{ssub:fund_distress}

We have available several measures of fund
distress: 1) fund profit and loss, 2) funds entering margin call state, and 3)
funds defaulting (when losses exceed capital). In the analysis that follows we
adopt funds entering default as our fund-level systemic risk metric. The advantage
of this metric is that it represents an extreme situation, more-so than merely
entering margin call state, and also represents a terminal state of the fund (funds
do not participate further in the simulation once they are in default). Using
profit and loss would require the specification of the value of loss necessary to describe a fund as distressed. By using the notion of default we automatically calibrate this threshold (the relationship between profit/loss and default is stated in Section \ref{ssub:banks}).

Following \citet{caccioli2014stability} we measure financial contagion by considering the fraction of funds that enter default in each simulation run. If at least $\gamma=5\%$ of funds are in default by the end of the simulation, we join \citet{caccioli2014stability} in considering a \emph{default cascade} to have occurred. We define the \emph{probability of contagion} as the fraction of Monte-Carlo trials (see Section \ref{sub:simulation_overview}) in which a default cascade occurs. Again following \citet{caccioli2014stability}, we define the \emph{extent of contagion} as the fraction of funds in default, conditional on a default cascade having occurred. Formally, the probability of contagion is given by

\begin{equation}
P_{\mathit{contagion}} = \frac{1}{M}\sum_{m=1}^{M}\mathbb{I}_{\mathit{cascade}}(m),
\end{equation}

where $M$ is the number of Monte-Carlo trials for a single location in the parameter space. $\mathbb{I}_{\mathit{cascade}}(m)$ is an indicator variable representing the occurrence of a default cascade during trial $m \in [1,M]$, and is defined as

\begin{equation}
\mathbb{I}_{\mathit{cascade}}(m) =
\begin{dcases}
1 & \mathrm{if} \enskip f_{\mathit{default}}(m) > \gamma, \\
0 & \mathrm{otherwise},
\end{dcases}
\end{equation}

where $\gamma$ is a scalar threshold controlling the fraction of funds that must default before we consider a default cascade to have occurred. $f_{\mathit{default}}(m)$, the fraction of funds entering default during Monte Carlo trial $m$, is given by

\begin{equation}
f_{\mathit{default}}(m) = \frac{1}{n_f}\sum_{i=1}^{n_f}\mathbb{I}_{\mathit{default}}(i, m),
\end{equation}

where $\mathbb{I}_{\mathit{default}}(i, m)$ is an indicator variable representing the default of fund $i$ in trial $m$, and is defined as

\begin{equation}
\mathbb{I}_{\mathit{default}}(i, m) =
\begin{dcases}
1 & \text{if fund}\enskip i\enskip\text{defaults in trial}\enskip m, \\
0 & \text{otherwise}.
\end{dcases}
\end{equation}

The extent of contagion, $\Omega_{\mathit{contagion}}$, is defined by \citet{caccioli2014stability} as the expected fraction of funds entering default, given a default cascade occurs. Formally, let us define $\mathcal{D}$, the set of Monte Carlo trials in which a default cascade occurs, such that $\mathcal{D} = \{m \in [1,M] : \mathbb{I}_{\mathit{cascade}}(m) = 1\}$, then we may write

\begin{equation}
\Omega_{\mathit{contagion}} = \frac{1}{|\mathcal{D}|}\sum_{m \in \mathcal{D}}f_{\mathit{default}}(m).
\end{equation}

Note that $\Omega_{\mathit{contagion}}$ is undefined if no default cascades occur, i.e. if $\mathbb{I}_{\mathit{cascade}}(m) = 0 \enskip \forall \enskip m \in [1, M]$, giving $|\mathcal{D}|=0$.

\subsubsection{Flash Crash Occurrence}
\label{sub:flash_crash_occurrence}

The second systemic risk notion concerns distress at assets, and here again we
have a choice of possible metrics. \citet{paddrik2012agent} look at the lowest price attained by simulated assets, which has the benefit of simplicity but lacks consideration of the time dimension so characteristic of flash crash phenomena. Although price falls are indeed material to our investigation, our
research questions are also concerned with the speed at which distress propagates,
and the temporal characteristics of sudden price movements are better-described
by metrics that are specifically designed to detect flash crashes. \citet{johnson2012financial} consider both price movement magnitude and timescale, but also require an unbroken sequence of trade prices in a given direction \citep[p.3]{johnson2012financial}:

\begin{displayquote}
``For a large price drop to qualify as an extreme event (i.e. black swan crash)
the stock price had to tick down at least ten times before ticking up
and the price change had to exceed 0.8\%. ... In order to explore timescales
which go beyond typical human reaction times, we
focus on black swans with durations less than 1500 milliseconds''
\end{displayquote}

We encountered difficulty calibrating \citet{johnson2012financial}'s method. In particular the metric was not found to be robust to simulation scaling (Section \ref{sub:simulation_overview}).

\citet{vuorenmaa2014agent} define a flash crash as a price fall of 10\% followed by recovery of at least 50\% of the size of the fall over an unspecified short timescale. \citet{jacob2016rock} use a similar metric based on a price fall of 5\% followed by an unspecified degree of reversion within a period of 30 minutes. The dependence on price reversion in each of these metrics was found to be problematic for our model where prices may not revert within the simulated timescale.

We define a flash crash as a price fall of 5\% over a rolling 5-minute window. No further constraints are applied. This metric was found to be robust to simulation scale, and to provide a better false positive rate than using the occurrence of volatility auctions (triggered by falls of 1.3\% in a single second, see Section \ref{ssub:assets}). Formally, let us define a variable, $\mathbb{I}_\mathit{FC}$, indicating the presence of a flash crash in asset $j \in [1,n_a]$ at time $t \in [1,T]$, as

\begin{equation}
\mathbb{I}_{\mathit{FC}}(j, t) =
\begin{dcases}
1 & \frac{(p_j^t - p_j^{t'})}{p_j^{t'}} < -0.05, \\
0 & \text{otherwise},
\end{dcases}
\end{equation}

where $p_j^{t'}$ is the price of asset $j$ at $\text{max}(t - 5 \enskip \text{minutes}, 0)$.

\subsubsection{Flash Crash Propagation Speed}
\label{ssub:flash_crash_propagation_speed}

In addition to considering the fraction of assets experiencing flash crashes
according to the above definition, we also consider the speed of flash crash
propagation through the system. Propagation speed is only well-defined where
the number of assets experiencing flash crashes is at least two. We measure
the time since the start of the simulation run at which assets experience their
first flash crash (individual assets may go on to experience several
flash crashes but we do not need to account for this in our propagation speed
metric). We then sort these time intervals in ascending order and take the
time difference between the first flash crash and the 80th percentile flash crash
as a measure of the time taken for flash crash propagation in the system.
We scale this time interval inversely according to the number of assets that experience
flash crashes, giving a resulting metric with units of flash-crashes per minute.

Let us define the set $\mathcal{F}^j$ of flash crashes experienced by asset $j$ such that $\mathcal{F}^j = \{t \in [1,T] \enskip : \enskip \mathbb{I}_\mathit{FC} (j,t) = 1 \}$,
where $T$ is the total number of simulated time steps, and $\mathbb{I}_\mathit{FC} (j,t)$ is the indicator variable defined above.

Let us further define a total order $\leq_j$ over $\mathcal{F}^j$, then we may label the elements $t^j_i \in \mathcal{F}^j$ with indices $i \in [1,|\mathcal{F}^j|]$ such that $t^j_1 \leq_j t^j_2 \leq_j ... \leq_j t^j_i \leq_j ... \leq_j t^j_{|\mathcal{F}^j|}$. We define the set $\mathcal{F}$ of first flash crashes experienced by each asset in a single Monte Carlo trial as

\begin{equation}
  \mathcal{F} = \bigcup_{j=1}^{n_a}\{ t^j_1 \}.
\end{equation}

We also define a total order $\leq$ over $\mathcal{F}$. We label the elements $t_i \in \mathcal{F}$ with indices $i \in [1,|\mathcal{F}|]$ such that $t_1 \leq t_2 \leq ... \leq t_i \leq ... \leq t_p \leq ... \leq t_{|\mathcal{F}|}$, where $p$ is the index of the $\zeta$th fractile of the ordered set, i.e. $p = \lfloor \zeta|\mathcal{F}| \rfloor$. We fix $\zeta=0.8$ in the present work. Finally, we define the speed $s$ of flash crash propagation by

\begin{equation}
  s = \frac{c\zeta(|\mathcal{F}|-1)}{(t_p-t_1)}, \quad |\mathcal{F}| > 1,
\end{equation}

where $c$ is a dimensionless constant that allows us to convert from units of $\text{(time steps)}^{-1}$ to $\text{(minutes)}^{-1}$. If $\delta t$ is the duration of one simulated time step in milliseconds, we have $c = 60,000\delta t^{-1}$.

An important feature of this metric is that it smooths out propagation speed for a given MC trial. We observed that propagation speed exhibits non-uniform temporal structure including bursts of activity. However, we also observed that the speed results are noisy and depend strongly on network topology (we return to this matter in Section \ref{sub:effect_of_network_topology}). The smoothing implicit in this metric therefore yields a more robust estimate of the propagation speed than naively using the maximum speed attained in a given trial.

\subsection{Simulation Protocol Summary}
\label{sub:simulation_overview}

Our simulation proceeds according to the following steps:

\begin{enumerate}
    \item \emph{Setup (meta):} Exogenous selection of parameter space region to explore and the number of stochastic \emph{Monte-Carlo} (MC) trials to perform for each parameter setting. MC instances dispatched to high-performance computing cluster.
    \item \emph{Setup (per-trial):} Construction of agents with stochastic parameters determined via pseudorandom number generation, including fund--asset network construction.
    \item \emph{Intraday model:} Simulate $T=100,000$ periods of market activity, each step lasting $\delta t = 50$ milliseconds.
    \item \emph{Single period:} Agents are selected for order placement according to a Poisson process. LOBs are updated in industry standard price-time priority.
    \item \emph{Shock:} After $T^*=\lfloor0.2T\rfloor$ periods to allow for the decay of initial transient effects, a distressed selling agent is introduced for a single asset. Observation of subsequent system dynamics and measurement of systemic risk.
    \item \emph{Distributional outputs:} Repeated stochastic MC trials facilitate the collection of distributional results.
\end{enumerate}

Due to the stochastic, path-dependent nature of the model (a common feature of agent-based models), it is necessary to perform repeated experimental trials in order to build statistical confidence in the observed systemic dynamics \citep{fagiolo2017validation,franke2012structural}. This approach, widely known as \emph{Monte-Carlo} (MC) sampling, induces a considerable computational burden. Our simulation protocol accommodates the computational cost in two ways --- firstly by making highly-parallel use of a distributed high-performance computation cluster, and secondly by introducing judicious granularity into the agent simulation\footnote{Performance of the LOB software component was also found to be critical. We extended the implementation at \\\url{https://github.com/ab24v07/PyLOB} accessed 2017-01-31, available under the MIT license and developed as part of \citet{booth2016algorithmic}.}. Facing similar computational constraints, \citet{paddrik2012agent} introduce granularity by scaling the number of trader agents in their simulation\footnote{\citet{hayes2014agent} provide a reference software implementation at \\\url{https://github.com/uva-financial-engineering/JinSup} accessed 2017-02-23, available under a permissive license.}. Table \ref{table:trader_agents} gives unscaled agent populations. Following \citet{paddrik2012agent} we divide these populations by a factor of 32 (we found our systemic risk metrics were robust to simulation at $\sfrac{1}{32}$ or $\sfrac{1}{4}$ scale). Since this scaling results in reduced LOB volume, it is also necessary to scale agent inventory limits by the same factor. Agent activity timescales and order placement sizes are not scaled. We adopt the same time granularity used in the flash crash model of \citet{vuorenmaa2014agent} and set $\delta t = 50\text{ms}$. It is possible that the Poisson arrival processes governing agent timing result in more than one agent being selected to place orders on a single LOB during a single time period. In such cases, agent activation order is randomised. Since agents are therefore unable to have perfect knowledge of the LOB at the exact time of their order placement, this randomisation places a lower-bound on agent trading latency. The most active agents (HFTs) arrive at the market according to a characteristic timescale of 350ms (see Table \ref{table:trader_agents}), about an order of magnitude slower than the minimal order placement latency supported in our simulation.

\subsection{Mathematical Notation}
\label{sec:parameters}

Table \ref{table:params} presents a summary of the parameters used in this paper. The sampled domain for each parameter is discussed at point of parameter usage in the results sections that follow.
Table \ref{table:symbols} summarises other non-parameter notation.

\begin{table}[h!]
\centering
\begin{tabular}{c l l l}
\toprule
Symbol                   & Name                                        & Theoretical Domain        & Sampled Domain \\
\midrule
$n_f$           & Number of funds                             & $\mathbb{N}^+$            & $\lbrack0,195\rbrack$ \\
$n_a$           & Number of assets                            & $\mathbb{N}^+$            & $\lbrack1,195\rbrack$ \\
$A_{ij}$        & Asset portfolio values ($n_f\times n_a$)    & $\mathbb{R}$              & $\lbrack0,\sim\$100mn\rbrack$ \\
$S_{ij}$        & Asset portfolio positions ($n_f\times n_a$) & $\mathbb{N}^+$            & $\lbrack0,\sim 100,000\rbrack$\\
$M$             & Number of Monte Carlo replicates            & $\mathbb{N}^+$            & $\lbrack1,100\rbrack$ \\
$T$             & Number of simulated time steps              & $\mathbb{N}^+$            & 100,000 \\
$\delta t$      & Time step duration / milliseconds           & $\mathbb{N}^+$            & 50 \\
$\rho$          & \erdos network density                      & $\lbrack0,1\rbrack$       & $\lbrack0,1\rbrack$ \\
$k$             & Network degree of fund nodes                & $\mathbb{N}$              & $\lbrack1,195\rbrack$ \\
$\sigma$        & Standard deviation of \erdos degrees        & $\lbrack0, \infty)$       & 0.001 \\
$\beta$         & Preferential attachment coefficient         & $\mathbb{R}$              & $\lbrack-7.5,7.5\rbrack$ \\
$\lambda^0$     & Initial fund leverage                       & $\mathbb{R}^+$            & $\lbrack1,20\rbrack$ \\
$C^0$           & Initial fund capital /\$millions            & $\mathbb{R}^+$            & $\lbrack0.25,5\rbrack$ \\
$\tau_c$        & Margin hysteresis critical value            & $\lbrack1, \infty)$       & $\lbrack1.001, 1.2\rbrack$\\
$\bm{\theta}$   & Parameter triple $(\lambda^0, \tau_c, C^0)$ & -                         & -\\
$\alpha$        & Fund portfolio uniformity coefficient       & $\lbrack0,1\rbrack$       & $\lbrack0,1\rbrack$ \\
$\eta$          & Distressed seller volume fraction           & $\mathbb{R}^+$            & $\lbrack0.05,0.18\rbrack$ \\
$\delta$        & Distressed seller timescale / s             & $\mathbb{R}^+$            & $\lbrack1,60\rbrack$ \\
$\gamma$        & Default cascade indicator threshold         & $(0,1\rbrack$             & 0.05 \\
$\zeta$         & Flash crash speed measurement percentile    & $\lbrack0,1\rbrack$       & 0.8 \\
\bottomrule

\end{tabular}
\caption{Parameters in the model. The sampled domain for each parameter is discussed at point of parameter usage in the results sections.}
\label{table:params}
\end{table}

\begin{table}[h]
\centering
\begin{tabular}{c l l l}
\toprule
Symbol                   & Name                                        & Theoretical Range         \\
\midrule
$V_i$           & Total investment value for fund $i$         & $\mathbb{R}^+$             \\
$L_i$           & Total loan value made to fund $i$           & $\mathbb{R}^+$             \\
$p_j^t$         & Price of $j$th asset at time $t$            & $\mathbb{R}^+$            \\
$Q_j^{\mathit{bid}}$       & Total shares of $j$th asset at best bid price       & $\mathbb{N}^+$  \\
$Q_j^{\mathit{ask}}$       & Total shares of $j$th asset at best ask price       & $\mathbb{N}^+$  \\
$P_{\mathit{contagion}}$ & Probability of contagion & $\lbrack0,1\rbrack$ \\
$\Omega_{\mathit{contagion}}$ & Extent of contagion & $\lbrack0,1\rbrack$ \\

\bottomrule

\end{tabular}
\caption{Other notation used in this paper.}
\label{table:symbols}
\end{table}

\section{Empirical Validation}
\label{sub:empirical_validation_results}

\begin{figure}
\centering
\makebox[\textwidth][c]
{
  \begin{tabular}{cc}
    \includegraphics{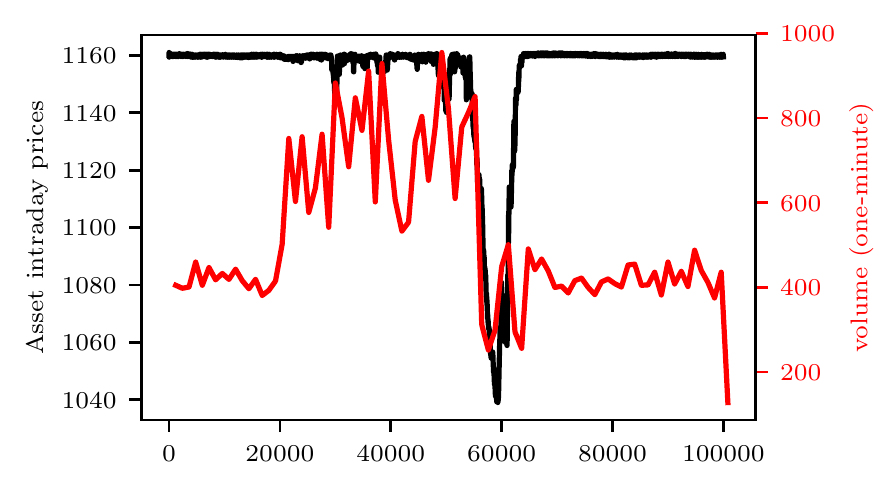} & \includegraphics{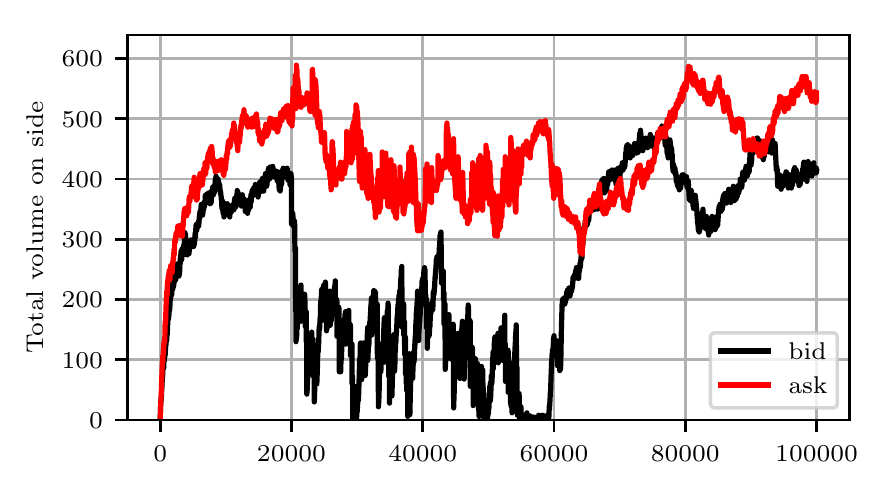} \\
    (a) & (b) \\ [7pt]
    \includegraphics{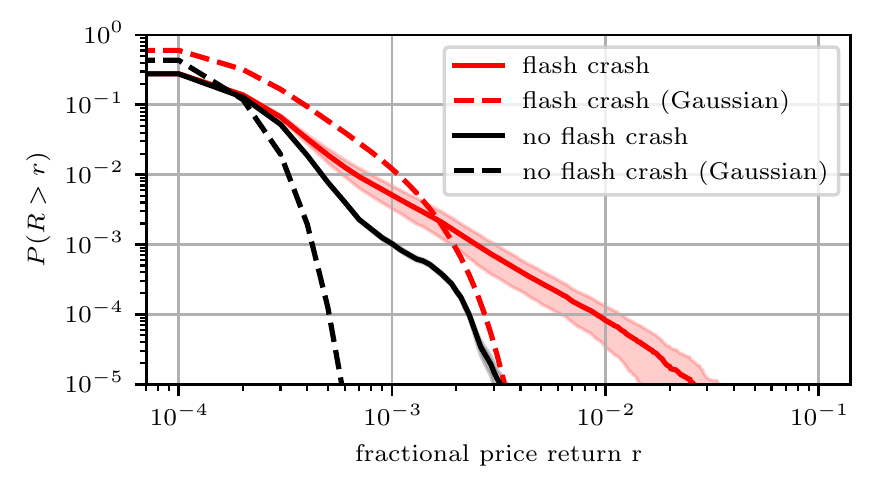} & \includegraphics{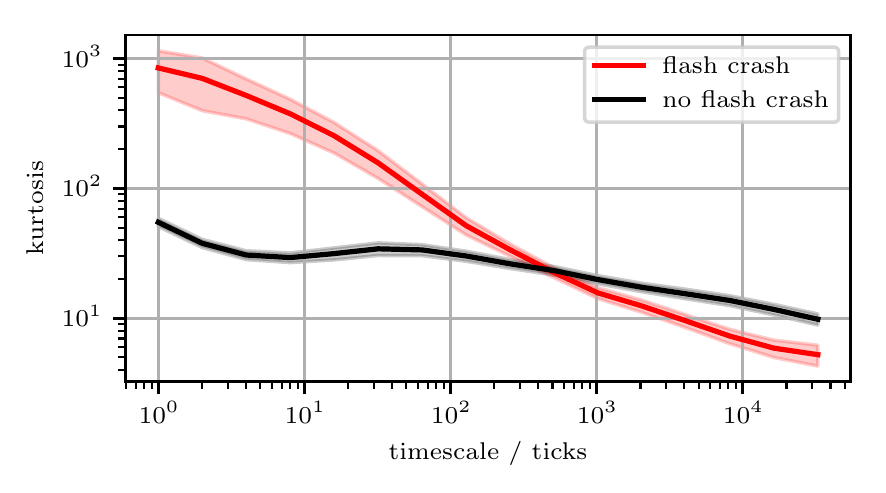} \\
    (c) & (d) \\ [7pt]
    \includegraphics{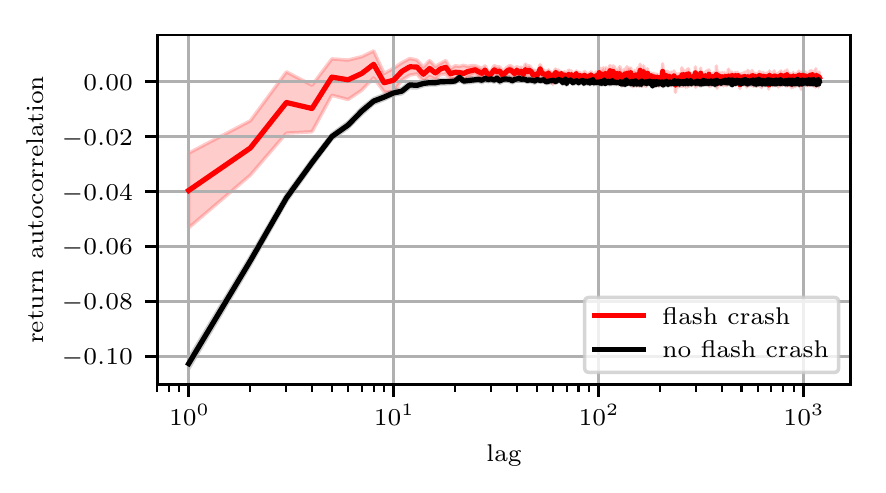} & \includegraphics{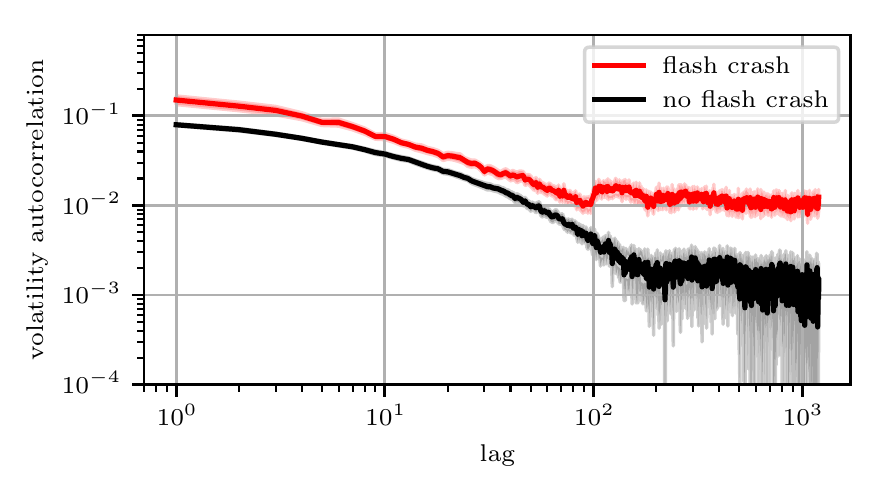} \\
    (e) & (f) \\ [7pt]
  \end{tabular}
}
\caption{\double \incolour Sample model output for a single asset. (a) price and volume
series demonstrating the occurrence of a flash crash event. (b) Total volume on bid (black)
and ask (red) sides of the LOB demonstrating the collapse of buy-side volume prior
to the flash crash. The simulation lasts for $T=100,000$ $\times$ $\delta t=50$ms simulated time steps, therefore represents
approximately 1.5 hours of market activity. (c) cumulative distribution function over price returns (log-log scale).
(d) return kurtosis measured over increasing time horizons (log-log scale).
(e) demonstrating lack of autocorrelation of price returns except at
very short horizons (x-axis log scale).
(f) demonstrating slowly-decaying autocorrelation of volatility
(log-log scale). For panels (c)-(f) we plot
the mean across 160 Monte-Carlo trials with 95\% confidence interval.}
\label{fig:stylised}
\end{figure}

Figure \ref{fig:stylised}(a) demonstrates the output from a single evaluation
of the ABM in which we simulate a single asset and a single fund. The simulation
comprises $T=100,000$ time steps each representing $\delta t=50$ms for a total of approximately
1.5 hours of simulated trading. The simulation was performed at $\sfrac{1}{4}$ scale following
the method established by \citet{paddrik2012agent}.

The system is shocked after an initialisation period
at $t=20,000$ by forcing the fund to liquidate all of its inventory.
The initial asset holdings of the shocked fund and its liquidation
behaviour are calibrated to generate equivalent
sell order placement activity to that exhibited by the distressed trader in the 6th
May 2010 Flash Crash \citep{kirilenko2017flash}. Figure
\ref{fig:stylised}(a) shows the price evolution of the system (black, left hand
axes) and the one-minute binned trade volume (red, right hand axes). We can see
the characteristic crash and recovery signature of a flash crash occurring around
$t=60,000$. Market volume and market volatility both increase while the
distressed selling is taking place and return to pre-crash levels after the
distressed selling has concluded. The zero-intelligence nature of the trader agents
precludes any temporally-extended response to volatility and so, unlike in the
real Flash Crash of 6th May 2010, the model does not capture the extended period
of volatility that followed the real crash \citep{kirilenko2017flash}.

Figure \ref{fig:stylised}(b) shows the total volume on the
LOB during the simulation disaggregated into bid (black) and ask (red) volumes.
We can see that the available bid volume collapses following the introduction of the
distressed selling agent at $t=20,000$. As discussed in Section \ref{ssub:traders},
the distressed selling agent places sell market orders which aggressively
remove liquidity from the bid side of the LOB. This figure clearly
shows the cause of the eventual flash crash --- just before $t=60,000$ there is
a brief period where buy-side liquidity collapses to almost zero. In such a regime,
the orderly provision of liquidity to the market has failed, and \emph{stub quotes} (orders placed far from prevailing prices that are not intended to transact)
may be executed. This is because market orders
do not have price limits and so trades can occur at arbitrarily low prices
\citep{cristelli2010liquidity}. Once the distressed
selling has completed after $t=60,000$, buy side volume returns to pre-distressed
levels, and fundamental buyer agents are able to restore the price to pre-crash
levels.

Figures \ref{fig:stylised}(c--f) demonstrate that the model reproduces several important
stylised facts of the market \citep{cont2001empirical, kirman2002microeconomic, chen2012agent}.
Figure \ref{fig:stylised}(c) demonstrates that the model produces heavy-tailed returns with
a power-law tail. We give the mean cumulative distribution function
across 160 Monte-Carlo trials with 95\% confidence intervals for trials
with (red) and without (black) flash crashes. For comparison, we also show Gaussian
return distributions with mean and variance matched to the observed results.
Whether a flash crash occurs or not, the returns are heavy tailed. The heavy-tailed
nature is much enhanced where flash crashes do indeed occur.

Figure \ref{fig:stylised}(d) demonstrates \emph{aggregational
Gaussianity} \citep{kirman2002microeconomic}, whereby returns taken over longer time horizons appear more Gaussian
and less heavy-tailed. Again, we show 95\% confidence intervals around the mean
across 160 Monte-Carlo trials disaggregated into trials with and without flash
crashes. We plot the kurtosis of the return distribution (a Gaussian distribution
has $kurtosis=3$). Both series converge towards the Gaussian kurtosis level as we
increase the measurement horizon over 10,000 simulated time-steps, and the decay is significantly
more pronounced for trials with a flash crash.

Figure \ref{fig:stylised}(e) demonstrates a lack of autocorrelation of price
returns except at very short intraday timescales, and Figure \ref{fig:stylised}(f) demonstrates
a long tail in the decay of autocorrelation of volatility (absolute return). Both
panels plot mean results across 160 Monte-Carlo trials with 95\% confidence intervals.
The slow decay of autocorrelation is markedly slower for trials exhibiting flash
crash phenomena. These results are consistent with empirically-observed stylised facts \citep{cont2001empirical}.

\citet{paddrik2012agent}'s model formulation requires all order placement by all
agents to be within ten \emph{ticks} (quantised minimum price increments on the LOB) of the current best price. Their model is
therefore precluded from generating the long tailed relative price stylised fact
reported by \citet{zovko2002power}. A consequence of that modelling decision
is that their model does not generate stub quotes. As a result, the \citet{paddrik2012agent} flash crash price gradient appears shallow
relative to an empirical flash crash or those generated by other models such as
\citet{vuorenmaa2014agent}. Our model relaxed the ten-tick assumption for the
small trader agents only, permitting them to place orders at up to one-thousand
ticks away from best. This resulted in an as-expected improvement in the model's
ability to reproduce the relative price stylised fact. Flash crashes generated under this relaxed assumption exhibit a steeper price gradient, in better agreement with empirical data (Figure \ref{fig:stylised}(a)).

For all of these simulations we held-fixed the distressed selling agent properties. Flash crash occurrence is, however, strongly dependent on the price impact
of the distressed selling agent's orders, which we would expect in turn to strongly
depend on order size and order placement rate \citep{paddrik2012agent}. We explore this matter in the next
section.

\FloatBarrier

\section{Trading Behaviour and Asset Price Shocks}
\label{sub:effect_of_asset_shock_magnitude}

The distressed selling agent in the Flash Crash of 6th May 2010 reportedly
placed market orders sized to not exceed 9\% of trailing one-minute market
volume \citep{kirilenko2017flash}. They furthermore executed a sequence of orders intended to achieve their
total desired sell volume over a period of 20 minutes. Although \citet{paddrik2012agent} and \citet{paddrik2015effects}
investigate sensitivity to the volume limit ($\eta$ in our notation), they do
not investigate sensitivity to the placement rate ($\delta$ in our notation). We
find that flash crash occurrence is critically sensitive to both parameters.

In this experimental scenario we retained the single-fund, single-asset structure
adopted in the previous section. The method of shock delivery via the complete
liquidation of the fund portfolio is also retained. We performed a combinatorial
evaluation over the parameter space $\eta \in \lbrack0.005,0.15\rbrack$ and
$\delta \in \lbrack1,60\rbrack$ sampling 77 parameter pairs.
For each parameter pair we performed 100 Monte-Carlo trials. Each trial was classified according to whether a flash crash was observed in the simulated asset price time-series. We define flash crash occurrence as a price fall exceeding 5\% in a period of 5 minutes (see Section \ref{sub:flash_crash_occurrence} for details regarding this metric).

\begin{figure}

\begin{subfigure}{\textwidth}
\centering
\makebox[\textwidth][c]{\includegraphics{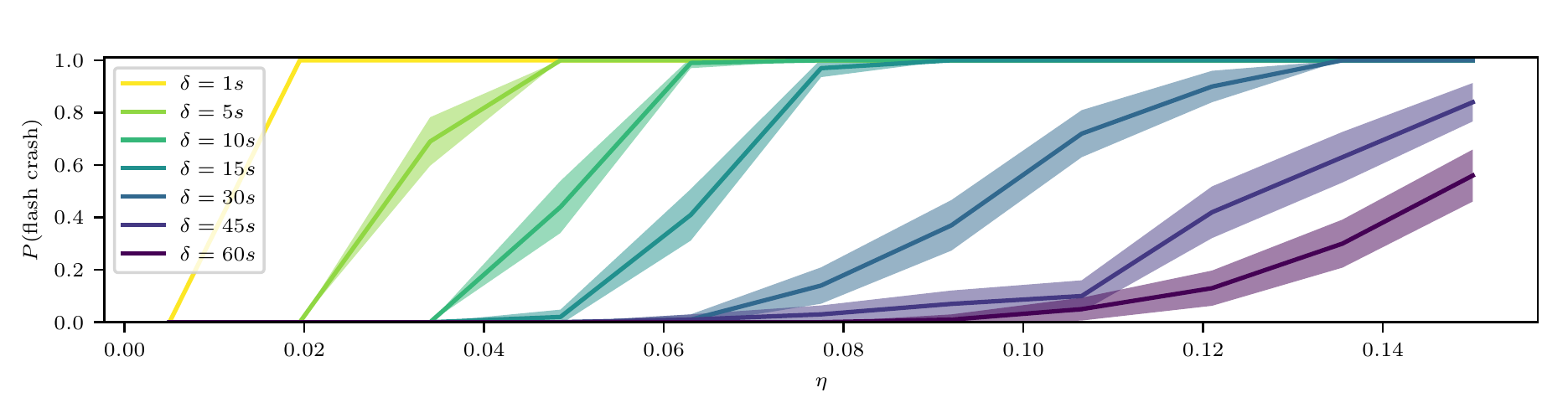}}
\caption{Probability of flash crash occurrence.}
\label{fig:distressed_seller_sensitivity}
\end{subfigure}

\begin{subfigure}{\textwidth}
\centering
\makebox[\textwidth][c]{\includegraphics{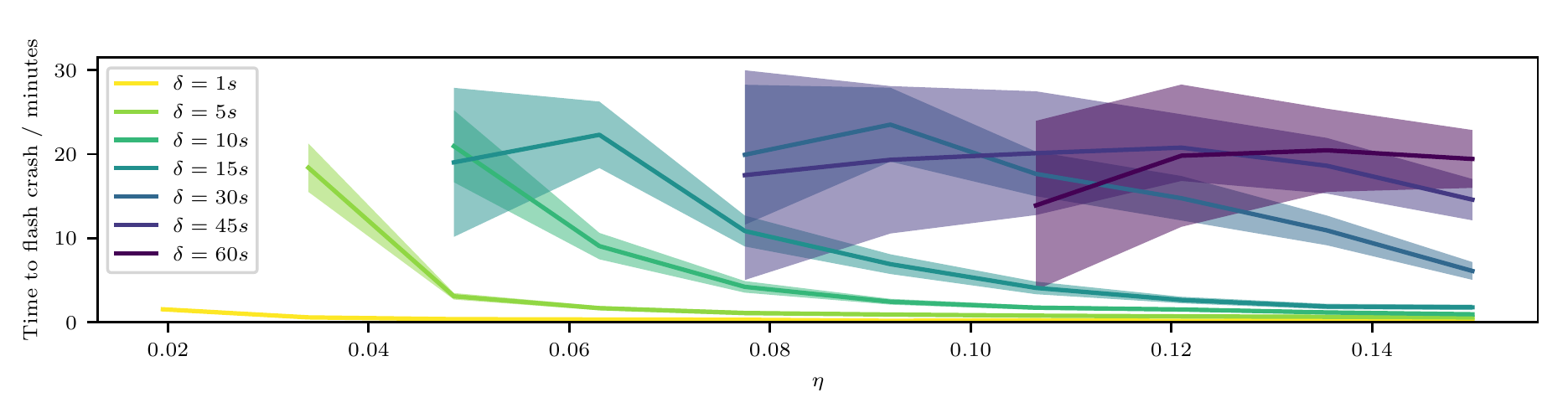}}
\caption{Time interval prior to flash crash occurrence following the onset of distressed selling.}
\label{fig:distressed_seller_sensitivity_onset}
\end{subfigure}

\caption{\double \incolour Single asset flash crash (a) occurrence and (b) time interval prior to occurrence as a function of distressed selling agent properties. $\eta$ refers to the order size limit as a fraction of rolling one-minute market volume. $\delta$ refers to the mean interval between sell order placements in seconds. We show the mean and 95\% confidence intervals sampled over 100 Monte Carlo trials.}

\end{figure}

The mean and 95\% confidence intervals for Monte Carlo results are plotted in Figure \ref{fig:distressed_seller_sensitivity}. What can be clearly seen in this figure is that flash crashes tend to occur with higher probability as the sizes of orders placed increase, and as the interval between order placements decreases. Both of these findings are consistent with the notion that a flash crash occurs when there is insufficient buy-side liquidity to meet elevated sell-side demand, possibly exacerbated by algorithmic traders withdrawing liquidity from the market. We find that, for any given order size limit in the regime studied, it is possible to reduce the probability of flash crash occurrence by increasing the interval between order placements. Figure \ref{fig:distressed_seller_sensitivity} also demonstrates that for a given order placement interval, $\delta$, there exists a critical value of the order size limit, $\eta$, below which flash crashes occur with negligible probability. Two traders placing equally-sized orders at equal timescales $\delta$ is exactly equivalent to a single trader placing orders with a timescale $\delta' = \sfrac{\delta}{2}$.

The distressed seller implicated in the Flash Crash of 6th May 2010 reportedly used a value of $\eta=0.09$ \citep{kirilenko2017flash}.
We calibrate this parameter setting in our model using this reported value, and retain this setting for the remainder of this paper. Although we are not able to calibrate the order placement rate in a similar manner, we select $\delta=10$s as a reasonable setting based on the results in Figure \ref{fig:distressed_seller_sensitivity}. Figure \ref{fig:distressed_seller_sensitivity_onset} implies that this choice of $\eta$ and $\delta$ is sufficient to induce a flash crash within approximately two minutes of the onset of distressed selling.

\FloatBarrier

\section{Leverage, Capital, and Margin Tolerance}
\label{sub:effect_of_leverage}

Previous sections have established the effectiveness of the model at reproducing
stylised facts of real financial markets, as well as its ability to exhibit flash
crashes. We now proceed to consider a full simulation of multiple connected
funds and assets, as opposed to the single-fund single-asset
simulations studied in previous sections.

In this section, we consider the response of the system to changes in fundamental
parameters related to margin tolerance $(\tau_c)$, leverage $(\lambda^0)$
and fund capital $(C^0)$. We find that the multi-asset system
exhibits phase changes for critical values of each of these parameters, moving
from a stable phase where it is unlikely for fund defaults and flash crashes to
occur, to an unstable phase where catastrophic collapse becomes highly likely.

Formal definitions of our systemic risk metrics for fund default cascades and asset flash crash occurrence are given in Section \ref{sub:systemic_risk_metrics}. In brief, we consider a default cascade to have occurred if more than $\gamma = 5\%$ of funds end the simulation in default. We further define the extent of a default cascade to be the fraction of funds in default, conditional on a default cascade having occurred. Both of these metrics follow the methodology of \citet{caccioli2014stability}. We also introduce our own metric for flash crash occurrence, which we define as an asset price fall of 5\% within a rolling five minute window.

\subsection{Experiment Parameters}
\label{sub:experiment_parameters}

For this experimental scenario we perform a combinatorial exploration over parameter triples
$\bm{\theta} (\lambda^0, \tau_c, C^0)$ for
$\lambda^0 \in \lbrack1,20\rbrack$, $\tau_c \in \lbrack1.002, 1.1\rbrack$
and $C^0 \in \lbrack0.1, 5.0\rbrack$ \$millions (210 total combinations).
The analysis performed by
\citet{ang2011hedge} suggested that real-world $\lambda^0$ typically lies within
$\lbrack1,20\rbrack$ for hedge funds trading equities. Since we do not have access to an independent
data source for our calibration, we adopt \citet{ang2011hedge}'s suggested domain. Analysis of the
6th May 2010 Flash Crash implied the size of sell orders sent by the distressed
trader totalled 75,000 \citep{kirilenko2017flash}. We sample
$C^0 \in \lbrack0.1, 5.0\rbrack$ such that in combination with $\lambda^0$
(see Section \ref{ssub:banks} for details), the positions held by funds extend from
the region whereby full liquidation of a single fund could not induce a flash crash
even if all of its holdings were allocated to a single asset, to the region
whereby liquidation of a single fund could indeed induce a flash crash. We do
not have access to empirical data on margin hysteresis $\tau_c$ and so here we
perform a logarithmic sampling across three orders of magnitude\footnote{i.e. we sample $\tau_c - 1 = \tau_c' \in \lbrack0.002, 0.1\rbrack$.} in order to determine
sensitivity to this parameter. In combination, these parameter domains are
sufficient to determine the boundary in $(\lambda^0, \tau_c, C^0)$
space separating stable from unstable behaviour. Each parameter triple, $\bm{\theta}$,
is evaluated over five Monte-Carlo trials\footnote{The low number is due to computational
constraints which can be improved in future studies.} resulting in a total of 1050 ABM evaluations. Only the pseudorandom
number seed is varied between Monte-Carlo trials.

For each evaluation we generate
a new set of fund-asset allocations according to the algorithm presented in
\ref{ssub:fund_asset_bipartite_network}. Allocations are stochastic with a fixed
diversification coefficient $\rho=0.5$ and no preferential attachment ($\beta=0$). The topology may therefore be represented as an \erdos random graph \citep{newman2003structure}
in which, on average, edges exist between each fund node and 50\% of the available asset nodes. We consider the effect of topology in the
next section. Finally, we fix the number of assets and funds at $n_a=n_f=50$. These
selections strike a balance between computational
tractability and providing fine enough grained distinction for metrics such as
flash crash propagation speed and extent. For the
present study, all funds
have the same level of capital $C^0$ and initial leverage $\lambda^0$.

The system is perturbed with an exogenous shock according to the same protocol as that presented in previous
sections. We introduce a distressed seller to a single asset (selected from a
uniform distribution across all assets that are held by at least one fund).
The distressed agent places a sequence of sell orders with a total quantity of shares that is calibrated
to trigger a flash crash in the shocked asset. No further exogenous interventions are performed, and further episodes of distressed selling by funds and asset flash crashes arise endogenously.

\subsection{Parameter Space Exploration}
\label{sub:parameter_space_exploration}

\begin{figure}
\centering
\makebox[\textwidth][c]{\includegraphics{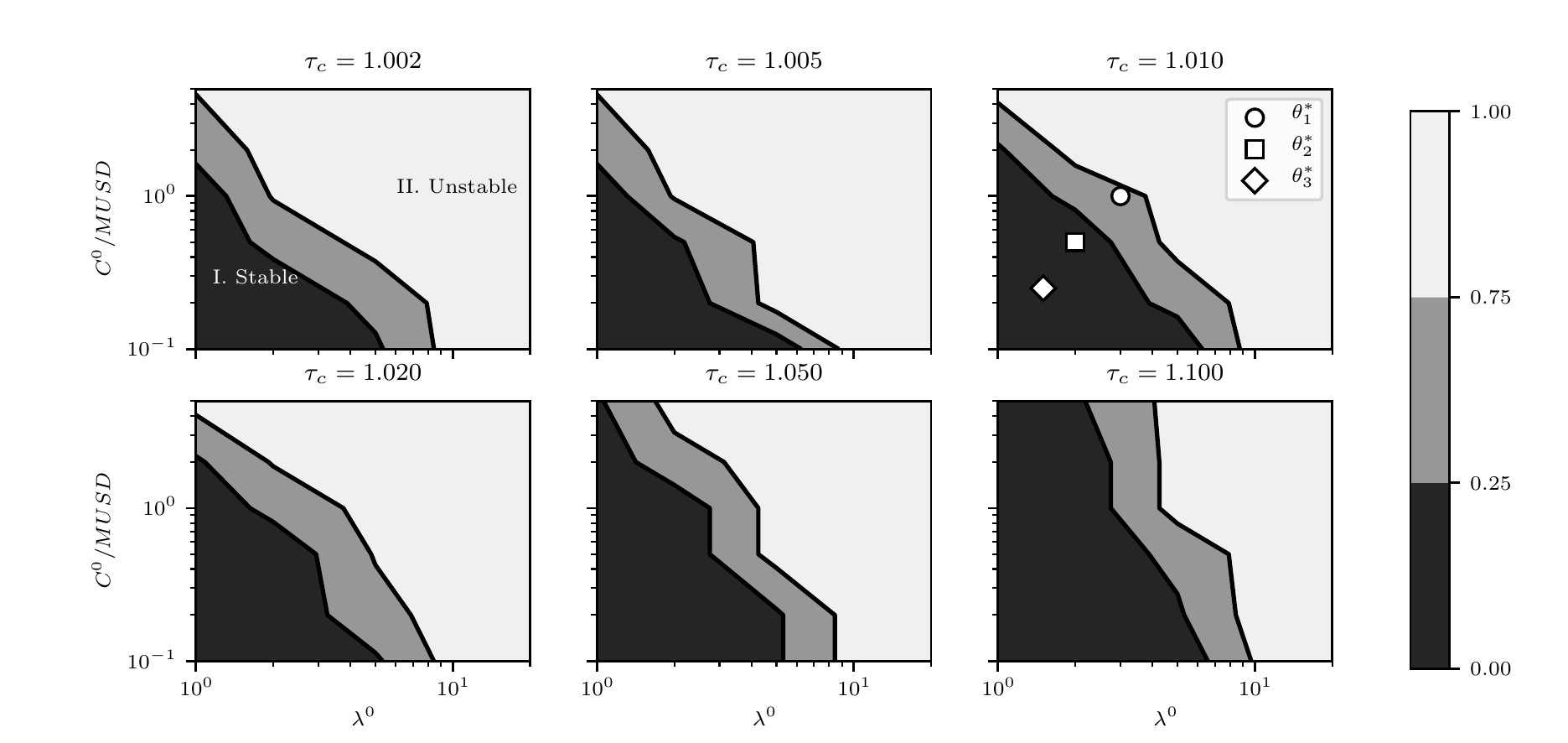}}
\caption{\double \incolour Probability of contagion as a function of leverage $(\lambda^0)$,
margin hysteresis $(\tau_c)$
and fund capital $(C^0)$. Plotted values are the mean across Monte-Carlo trials as a
function of $\lambda^0$ and $C^0$. $\tau_c$ increases from panel to panel thus
we present slices of the full three-dimensional parameter space at fixed values
of $\tau_c$. Contour lines are drawn to denote fractional default at the
25\% and 75\% levels. In region I most funds
avoid default, but in region II it is likely that all funds end up in default.
In the upper-right panel we label three points in $(\lambda^0, \tau_c, C^0)$ parameter space $\bm{\theta^\star_1}, \bm{\theta^\star_3}, \text{and} \bm{\theta^\star_3}$. We refer to these in detail in Section \ref{sub:effect_of_network_topology}.}
\label{fig:leverage_grid}
\end{figure}

Figure \ref{fig:leverage_grid} shows the fraction of funds entering default
(mean across Monte-Carlo trials) as we perform the combinatorial search. $\tau_c$ is
increased from panel to panel thus we present slices of the full three-dimensional
parameter space at fixed values of $\tau_c$. It is clearly seen from this figure
that a phase-change exists in leverage-capital space, whereby the likelihood of
funds entering default increases from approximately zero (region I) to approximately
100\% (region II). This result is consistent with expectations because a larger value of leverage $\lambda^0$ magnifies fund losses in response to a given asset price move, making margin calls more likely. This effect is directly mitigated by increasing margin tolerance $\tau_c$. Conditional on a margin call being in force, larger fund capital $C^0$ means that a greater quantity of shares will need to be sold by the fund, which therefore implies a greater price impact when that fund elects to sell, propagating greater losses to other funds that hold positions in the same assets. Systemic sensitivity to the topological properties of overlapping (crowded) asset portfolios are explored in detail in Section \ref{sub:effect_of_network_topology}.

\begin{figure}

\begin{subfigure}{\textwidth}
\centering
\makebox[\textwidth][c]{\includegraphics{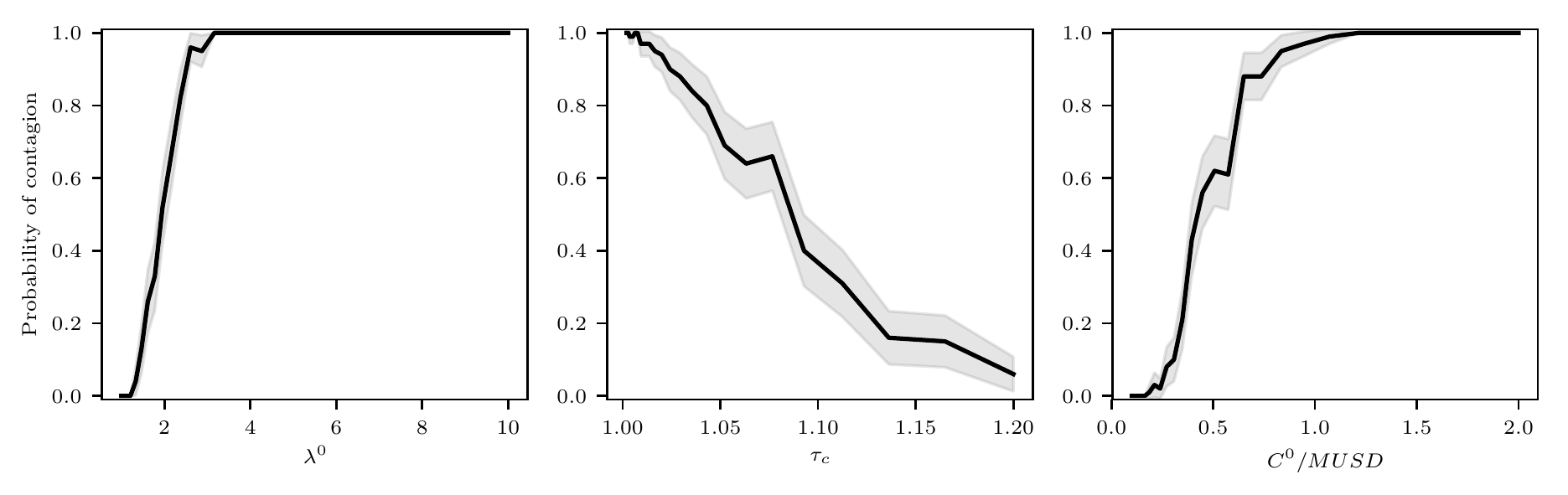}}
\caption{Probability of contagion as we systematically adjust one variable,
holding the other two fixed.}
\label{fig:leverage_line_search}
\end{subfigure}

\begin{subfigure}{\textwidth}
\centering
\makebox[\textwidth][c]{\includegraphics{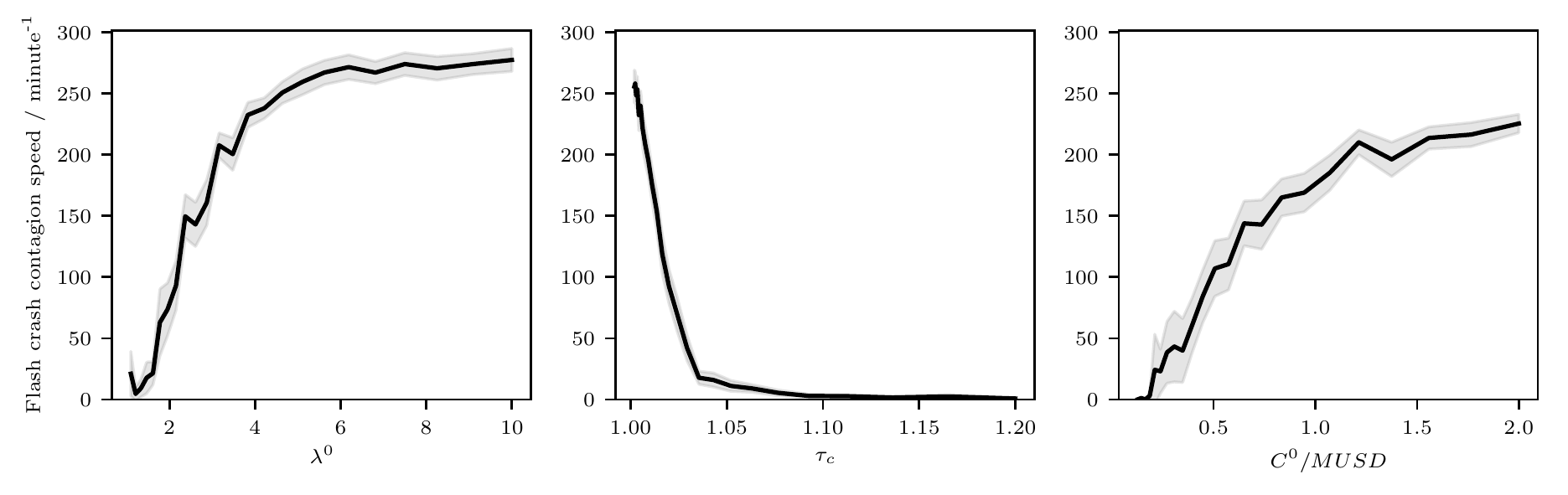}}
\caption{Flash crash propagation speed as we systematically adjust one variable, holding the other two fixed.}
\label{fig:leverage_line_search_fc}
\end{subfigure}

\begin{subfigure}{\textwidth}
\centering
\makebox[\textwidth][c]{\includegraphics{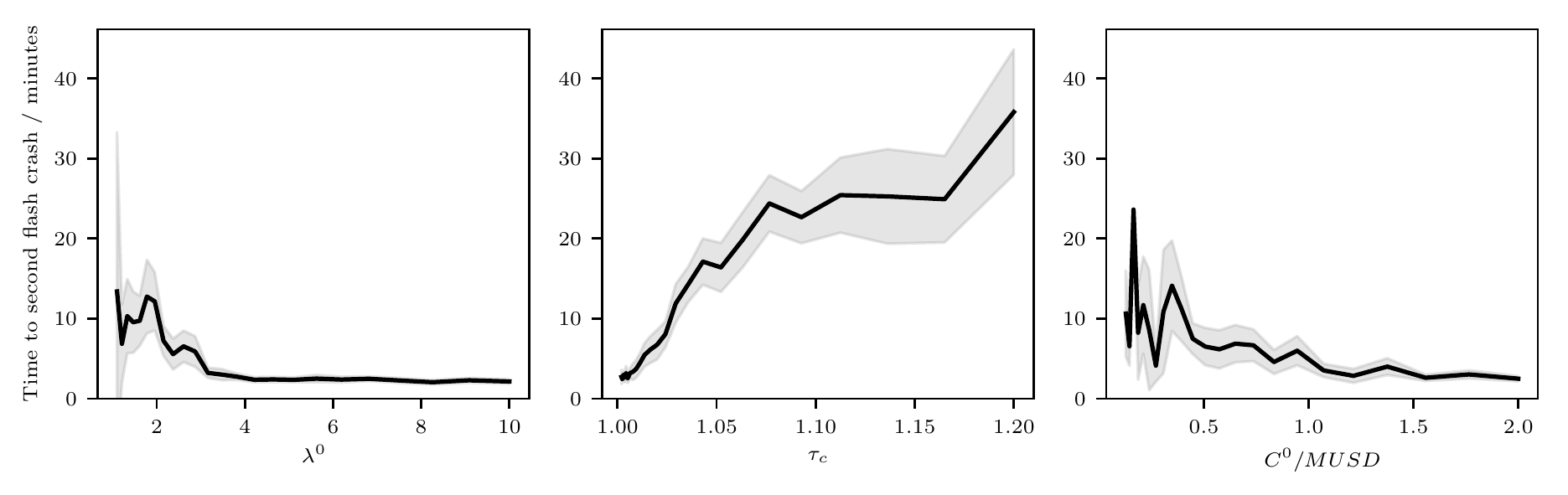}}
\caption{Time taken for the second flash crash to occur following the introduction of the distressed selling agent, as we systematically adjust one variable, holding the other two fixed.}
\label{fig:leverage_line_search_onset}
\end{subfigure}

\caption{\double \incolour Exploring financial stability as we systematically adjust one agent parameter,
holding the other two fixed. The exogenous variables are Left panels: leverage $\lambda^0$, Middle panels: margin hysteresis $\tau_c$, Right panels: capital $C^0$ (units millions of US dollars, \emph{MUSD}).
We plot the mean across 100 Monte-Carlo trials,
along with 95\% confidence intervals. The intersection in parameter space is at
$\bm{\theta^\star_1}(\lambda^0=3, \tau_c=1.01, C^0=1)$. The results demonstrate
phase changes for each of the three parameters from stable to unstable behaviour.}

\end{figure}

Although it is not easy
to see in Figure \ref{fig:leverage_grid}, a phase change also occurs over the domain
of $\tau_c$. In order
to make this more apparent, we performed linear sensitivity analyses of the three variables in
isolation, holding the other two variables fixed. The three sampled lines intersect at
$\bm{\theta^\star_1} = \bm{\theta}(\lambda^0=3, \tau_c=1.01, C^0=1)$. Figure
\ref{fig:leverage_line_search} plots the probability of contagion
for these linear explorations, showing 95\% confidence intervals
around the mean. This figure clearly demonstrates the phase changes in all three
variables under consideration. The change from stability to instability occurs
as leverage and capital increase and as margin tolerance decreases. We further observed that the extent of contagion was always 100\%, conditional on the occurrence of a default cascade. This result adds evidence in support of existing studies. Similarly to \citet{huang2013cascading},
we find that increasing margin tolerance ($\tau_c$) reduces the likelihood of systemic failure, and also that increasing price impact correspondingly increases systemic risk. In our model, price
impact is a function of the distressed liquidity entering the market which, by
construction, is an increasing function of leverage ($\lambda^0$) and capital ($C^0$).

Figure \ref{fig:leverage_line_search_fc} plots the speed of flash
crash propagation for the linear explorations. As discussed in Section \ref{ssub:flash_crash_propagation_speed},
the speed we report here is inversely related to the time taken for 80\% of subsequent
flash crashes to occur, following the observation of the first flash crash.
These results
demonstrate that flash crash propagation speed is an increasing concave function of leverage and capital, and
a decreasing convex function of margin hysteresis, similarly to the probability of contagion result. Margin hysteresis has a dramatic damping effect on flash crash propagation speed, reducing propagation speeds close to zero for $\tau_c>1.03$. These results are straightforward to explain. Increasing leverage makes funds more sensitive to price changes at an asset, making funds more likely to commence distressed selling for a given asset price move, and hence more readily propagating price impact between assets. Increased capital means that, conditional on distressed selling taking place, a fund will sell a larger quantity of shares, resulting in steeper price impact in assets that are being sold. This in turn means that the conditions satisfying our flash crash metric (a fall of 5\% within a rolling 5 minute window) will be satisfied sooner than under lower capital conditions. The effect of increasing margin hysteresis is predominantly in opposition to the effect of increasing leverage. However, conditional on flash crashes occurring under high margin tolerance, the total quantity of shares to be unwound will be higher than when tolerance is small. This leads to a mixed effect of the margin tolerance parameter.

Finally, Figure \ref{fig:leverage_line_search_onset} illustrates the time taken for the second flash crash to occur following the introduction of an exogenous shock to the system. The first flash crash typically occurs in the asset receiving the exogenous shock, so by measuring the time of the second flash crash we probe the systemic nature of flash crash propagation across the fund--asset network. These results are again consistent with expectations and demonstrate that higher values of $\lambda^0$ and $C^0$ destabilise the system, causing flash crashes to occur more rapidly following commencement of distressed agent selling. Increasing the margin tolerance $\tau_c$ stabilises the system, leading to larger intervals between shock delivery and flash crash onset.

Our model
continues to support the conclusions of models with a similarly-structured macroscopic
network component. Our work goes further
by allowing detailed consideration of distress propagation speeds and of the
intraday mechanism by which contagion occurs. Price impact in real markets does
not occur instantaneously, rather it manifests across longer time periods while trading strategies
are in operation. The realistic calibration of the microstructure component of our model allowed us to show that the market quickly reacts to initial distress, leading to contagious spreading of flash crashes across the network in a matter of minutes. This short time interval places an upper bound on the  window for the deployment of \emph{ex post} regulatory interventions. Our results demonstrate that increasing margin tolerance provides the most effective route to increasing the size of the \emph{ex post} intervention window. However, the short timescale associated with this window across large regions of the parameter space suggests that \emph{ex ante} precautions may have higher efficacy than \emph{ex post} alternatives.

\FloatBarrier

\section{Bipartite Network Topology}
\label{sub:effect_of_network_topology}

Previous works
such as \citet{acemoglu2015systemic} and \citet{gai2010contagion}
have established the importance of network topology when considering the systemic
risk associated with financial distress propagation. \citet{gai2010contagion}
introduce the notion of robust-yet-fragile networks in which higher network connectivity
is advantageous for stability in the presence of small shocks as banks are able
to share losses and avoid default. On the other hand, high interconnectedness
becomes problematic during high shock conditions by providing a large number of channels
for distress propagation. \citet{acemoglu2015systemic} similarly argue that in
low-shock regimes, more highly-connected networks are more stable. In this section we explore the extent to which these findings apply to a joint micro-macro model of systemic risk due to overlapping portfolios.

In Section \ref{sub:effect_of_leverage} we represented the fund-asset allocation network as an \erdos random graph \citep{newman2003structure}. We held fixed both the topology and the graph density which
was characterised by the fund's average asset diversification $\rho=0.5$. We now discuss systemic risk when network density and topology are varied.

\subsection{Experiment Parameters}
\label{sub:experiment_parameters}

Using the bipartite network generation algorithm, we construct networks in which
funds exhibit varying degrees of asset diversification (controlled by parameter $\rho$) and also varying
degrees of preferential asset attachment (controlled by parameter $\beta$).

As in Section \ref{sub:effect_of_leverage} where we looked at leverage dependence, we continue with a 50-asset, 50-fund model\footnote{The effect of variable network size was found to be similar to the effect of diversification parameter $\rho$ and so we choose to frame our discussion in terms of $\rho$ for the sake of brevity.}. We perform
a combinatorial set of evaluations for $\rho \in \lbrack0,1\rbrack$ and
$\beta\in \lbrack-7.5,-2,0,2,7.5\rbrack$ yielding a total of 35 parameter pairs. Each pair was
evaluated over 100 Monte-Carlo trials in which only the pseudorandom number generator seed varied between
simulation instances.

Based on the results from Section \ref{sub:effect_of_leverage}
we chose three fixed points in $(\lambda^0, \tau_c, C^0)$ space. These locations are depicted in the upper-right panel of Figure \ref{fig:leverage_grid} and are presented in Table \ref{table:theta_locations}. We performed a total of 10,500 ABM evaluations, each consisting of $T=100,000$ time steps, each step with a simulated resolution of $\delta t=50$ms.

As in previous sections, we introduce an exogenous shock to a single asset in the form of a distressed selling agent, calibrated to place a sequence of orders sufficient to trigger a flash crash in the selected asset. We select the asset to shock uniformly from the set of assets that are held by at least one fund. In this way we avoid shocking assets that are disconnected from the network. Formally, let us define the set of connected assets $\mathcal{A} = \{j \enskip : \enskip \sum_{i=1}^{n_f}A_{ij} > 0, \enskip j \in \lbrack1, n_a\rbrack, j \in \mathbb{N}^+\}$, then the probability of selecting an asset $j' \in \mathcal{A}$ as the single asset to receive the exogenous shock is equal to $\sfrac{1}{|\mathcal{A}|}.$\footnote{We further require that if more that 80\% of funds are connected in a giant network component, then the shock must be delivered to an asset that is also connected to the same component. This is to reduce the impact of spurious outliers and noise in the results and does not affect our conclusions.}

\begin{table}
\centering
\begin{tabular}{clllll}
  \toprule
  & $\lambda^0$ & $\tau_c$ & $C^0$ & leverage regime & stability\textdagger \\
  \midrule
  $\bm{\theta^*_1}$ & 3 & 1.01 &  1  &  high & typically unstable\\
  $\bm{\theta^*_2}$ & 2 & 1.01 &  0.5   &  medium & moderately stable\\
  $\bm{\theta^*_3}$ & 1.5 & 1.01 &  0.25  &  low & typically stable\\
  \bottomrule
\end{tabular}
\caption{Locations in $(\lambda^0, \tau_c, C^0)$ space which are fixed
during topology experiments. These points are depicted in Figure \ref{fig:leverage_grid}. \textdagger We report the stability at $\rho=0.5, \beta=0, n_f = n_a = 50$, which is consistent with parameter selection in Section \ref{sub:effect_of_leverage}.}
\label{table:theta_locations}
\end{table}

\subsection{Probability and Extent of Contagion}
\label{sub:probability_and_extent_of_contagion}

\begin{figure}
\centering
\makebox[\textwidth][c]
{
  \begin{tabular}{cc}
    \includegraphics{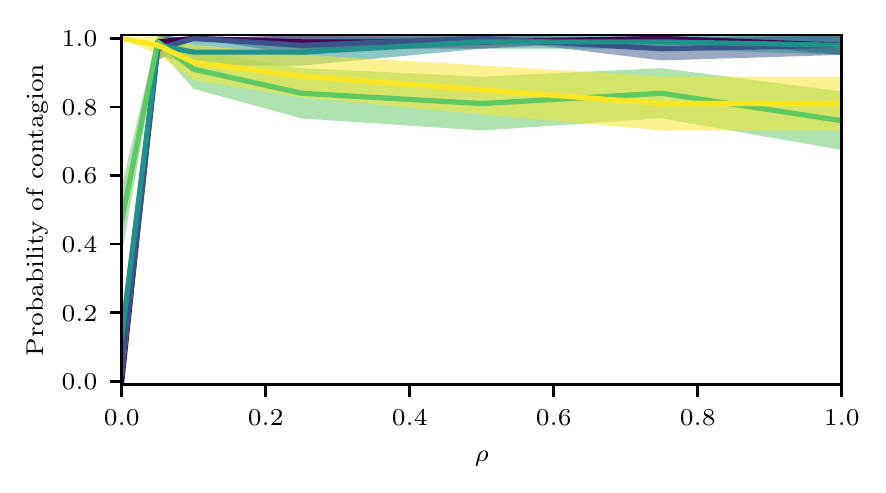} & \includegraphics{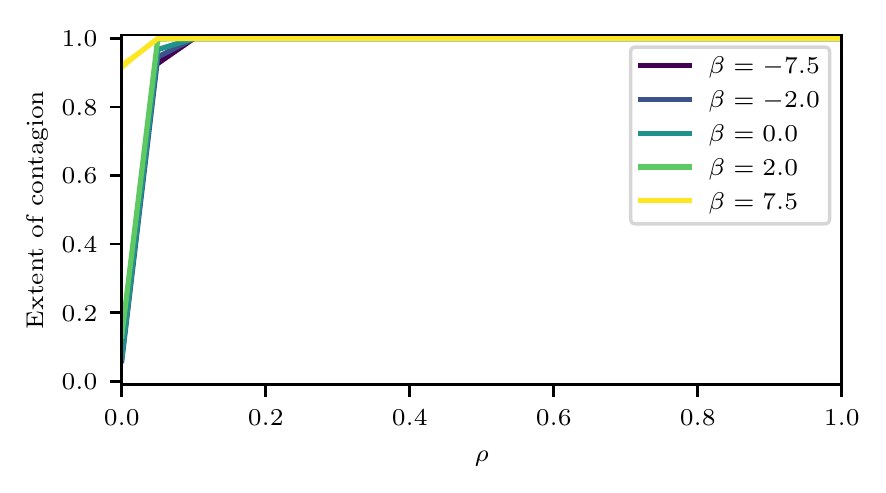} \\
    (a) $\bm{\theta^*_1}$ (high leverage) & (b) $\bm{\theta^*_1}$ (high leverage) \\ [7pt]
    \includegraphics{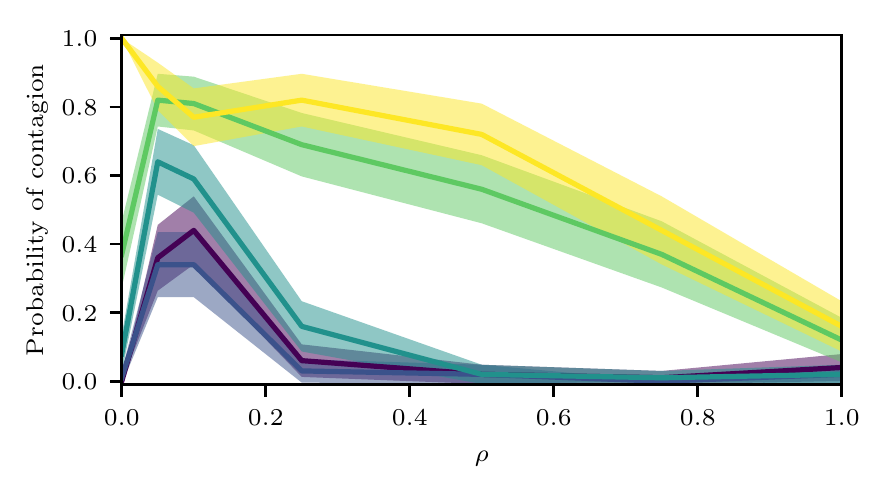} & \includegraphics{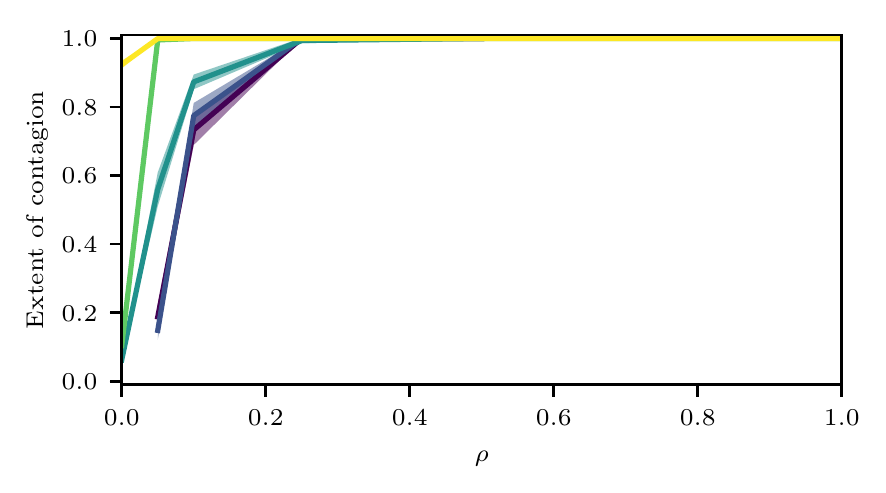} \\
    (c) $\bm{\theta^*_2}$ (medium leverage) & (d) $\bm{\theta^*_2}$ (medium leverage) \\ [7pt]
    \includegraphics{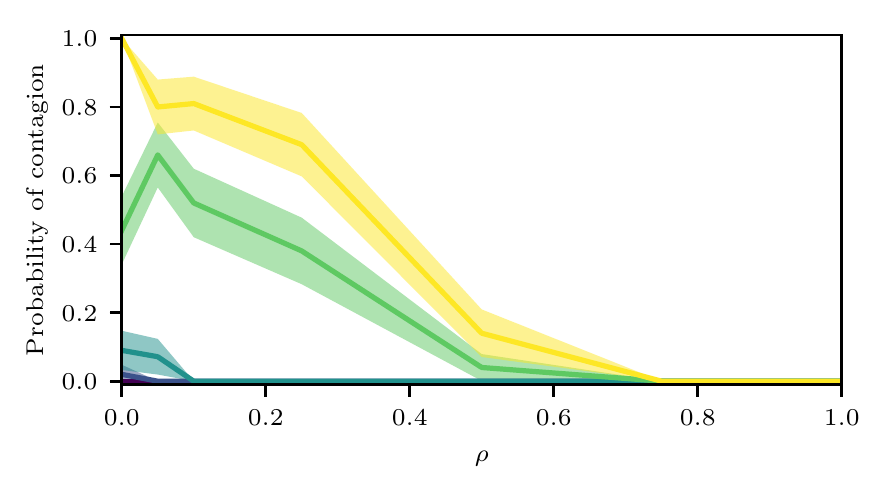} & \includegraphics{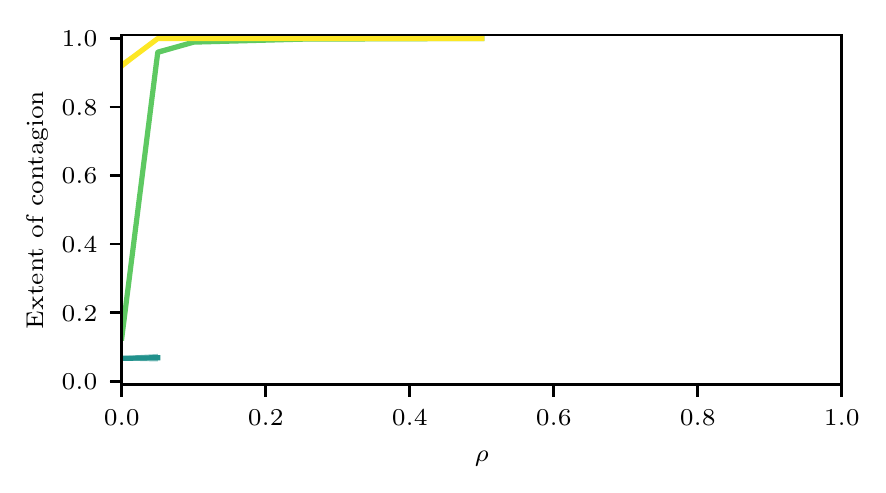} \\
    (e) $\bm{\theta^*_3}$ (low leverage) & (f) $\bm{\theta^*_3}$ (low leverage) \\ [7pt]
  \end{tabular}
}
\caption{\double \incolour Probability (a,c,e) and extent (b,d,f) of a default cascade as a function
of portfolio diversification ($\rho$) and crowding ($\beta$) parameters. The
analysis is repeated for three leverage regimes denoted $\{\bm{\theta^*_1}, \bm{\theta^*_2}$ and $\bm{\theta^*_3}\}$
in decreasing order of magnitude.}
\label{fig:crowding_contagion}
\end{figure}

Figure \ref{fig:crowding_contagion} presents the probability (left column) and extent (right column) of contagion as we vary network diversification ($\rho$) and crowding parameter ($\beta$). We present results for three leverage scenarios: high leverage ($\bm{\theta^*_1}$, top row), medium leverage ($\bm{\theta^*_2}$, middle row) and low leverage ($\bm{\theta^*_3}$, bottom row). We find that systemic risk is strongly dependent on each of these system parameters.

Figure \ref{fig:crowding_contagion}(a) shows that default cascades occur with high probability for all but the smallest values of diversification parameter $\rho$. When $\rho$ is less than 3\%, funds invest in a single asset\footnote{Recall $k_{\mathit{fund}} \approx max(\rho n_a, 1)$, and $n_a=50$ in the present experiment.}. At low diversification, we find that the system is most stable for dispersed configurations ($\beta < 0$). This corresponds to topologies in which each fund invests in a different asset, and so distressed selling by one fund cannot affect the portfolios of other funds (see Table \ref{table:connected_components}). \citet{chen2014asset} find a similar result for their purely analytic model. When diversification is low, we also observe that crowded systems ($\beta > 1$) are the least stable. This corresponds to a topology in which all funds invest in the same single asset. This has two implications --- first that a falling price of this asset will simultaneously affect all funds, and second that distressed selling by any fund will necessarily affect the single jointly-held asset. Both factors contribute to increased price impact at the jointly-held asset and hence flash crash propagation in the system. As diversification increases beyond $\rho=0.03$, the system becomes less stable for all crowding regimes. In high leverage scenarios, we therefore find that increasing diversification does not improve systemic stability. Rather, increased diversification allows contagion to spread freely to more of the assets and funds in the network. \citet{acemoglu2015systemic} find a somewhat analogous effect when comparing the size of shocks delivered to their system. However, a surprising result emerges as crowding ($\beta$) increases. It is not obvious what mechanism acts to increase system stability as a function of higher portfolio overlap for moderate and high levels of diversification in this high leverage regime. This finding is not consistent with existing literature, and we return to consider it in detail in Section \ref{sub:effect_of_asset_allocation_distribution}.

Figure \ref{fig:crowding_contagion}(c) presents the probability of contagion in a moderate leverage regime. At very low diversification we again observe that the system is most stable for dispersed configurations ($\beta < 0$). We may attribute this finding to the network topology which consists of disjoint sub-graphs in which each fund invests in a single asset, and each asset is held by exactly one fund. We find that cascades cannot propagate as the network is too sparsely connected, similar to the findings by \citet{caccioli2014stability}. As diversification increases to 5--10\% we observe a marked reduction in system stability for all crowding regimes. At this level of diversification the network is sufficiently connected for distress to propagate. However, as diversification increases further we find that systemic risk is reduced which is again consistent with \citet{caccioli2014stability}, who found that increased diversification means that distress at a fund will be spread between a larger number of assets and so have a smaller effect on any one asset. In the moderate leverage regime, we find that distress at a fund is less likely to cause a flash crash at a given asset. This stands in contrast to the high leverage scenario discussed previously where even fully diversified funds can still have sufficient price impact to trigger flash crash cascades. Highly-crowded regimes ($\beta>0$) also benefit from increased diversification, however such regimes are always less stable than non-crowded regimes. This result is to be expected: when portfolios are strongly-overlapping, distress at a single asset is felt by a large fraction of funds. Distress at a single fund is therefore more readily felt by other funds in crowded regimes.

Figure \ref{fig:crowding_contagion}(e) presents results when leverage is further decreased. We find that crowded regimes remain the least stable, regardless of the level of portfolio diversification. However, the figure shows that the benefits of portfolio diversification manifest at lower levels of diversification parameter $\rho$. In particular, no default cascades are observed for $\rho \geqslant 0.75$. When funds utilise low leverage, they are relatively insensitive to price movements of the assets in their portfolios and so even modest levels of diversification are sufficient to avoid the need for funds to engage in distressed selling. In the low leverage regime our results are consistent with those of \citet{chen2014asset} who find that systemic risk is minimised when all funds hold the ``market portfolio'', which in our notation corresponds to maximum diversification ($\rho=1$).

\begin{table}
\centering
\begin{tabular}{rrc}
\toprule

$\beta$ & $\langle n_f^C \rangle$ & 95\% interval\\

\midrule

-7.5 & 1.52 & 1.42 -- 1.62 \\
-2.0 & 2.56 & 2.45 -- 2.67 \\
 0.0 & 3.80 & 3.64 -- 3.96 \\
 2.0 & 11.2 & 10.6 -- 11.7 \\
 7.5 & 45.9 & 45.1 -- 46.7 \\

\bottomrule
\end{tabular}
\caption{Presenting the mean and 95\% confidence interval of $n_f^C$, the number of funds in the largest connected component of the $n_f = n_a = 50$ fund--asset network when $\rho \to 0$, i.e. $k_{\mathit{fund}} = 1$, for different choices of crowding parameter $\beta$. Statistics are generated from an ensemble of 100 random networks.}
\label{table:connected_components}
\end{table}

Figures \ref{fig:crowding_contagion}(b), \ref{fig:crowding_contagion}(d) and \ref{fig:crowding_contagion}(f) demonstrate that when default cascades are observed, they typically affect the entire network. That is, the extent of contagion is 100\% for anything other than very low levels of portfolio diversification. In the moderate and low leverage regimes the system is ``robust-yet-fragile'' \citep{gai2010contagion} for higher levels of diversification. Although default cascades are unlikely in such regimes, when they do occur the distress propagation is extensive. In each leverage regime, at low diversification the extent of contagion is highest for more highly-crowded systems. This effect is most pronounced in the moderate leverage regime. In the low leverage regime we did not observe cascades for many parameter combinations and so the (conditional) extent of contagion is undefined (this results in partial series, particularly noticeable in figure \ref{fig:crowding_contagion}(f)). When diversification is low, dispersed network topologies can become disconnected. Contagion is limited to the connected component (sub-graph) containing the exogenously-shocked asset, resulting in an extent of contagion of less than 100\%.

The presence of ``robust-yet-fragile'' behaviour creates a problem for policy-makers \citep{gai2010contagion}. Although systemic failure can be made less likely by encouraging higher portfolio diversification at funds, if a cycle of deleveraging takes hold, enhanced diversification means that financial distress spreads more extensively through the network of funds and assets, resulting in increased financial damage to the fund sector overall. In the next section we consider the dynamic nature of the spread of distress and investigate the speed of flash crash propagation as a function of network topology.

\subsection{Flash Crash Propagation Speed}
\label{sub:flash_crash_propagation_speed}

\begin{figure}
\centering
\makebox[\textwidth][c]
{
  \begin{tabular}{ccc}
    \includegraphics[width=63mm]{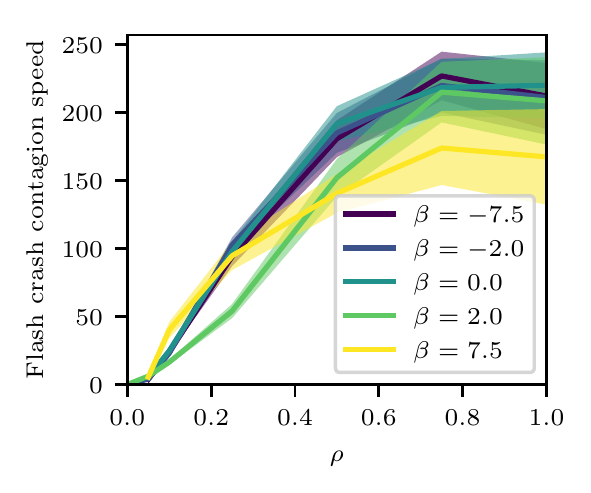} & \includegraphics[width=63mm]{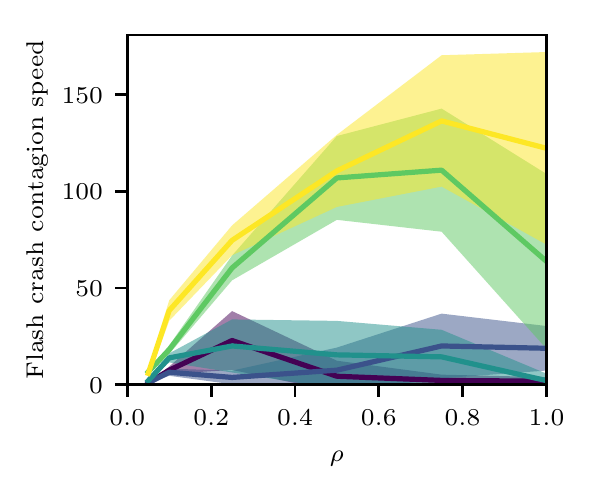} & \includegraphics[width=63mm]{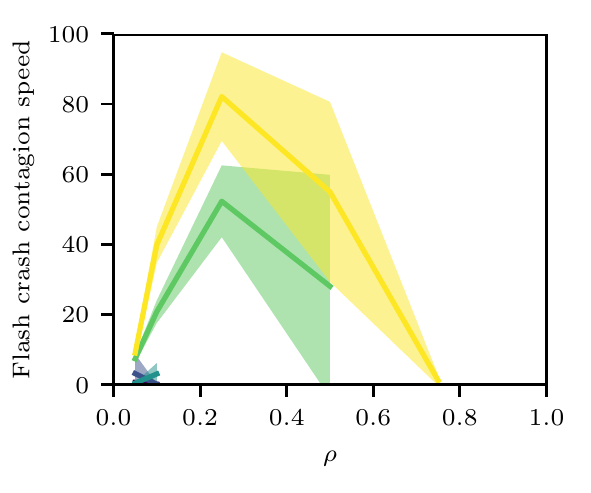} \\
    (a) $\bm{\theta^*_1}$ (high leverage) & (b) $\bm{\theta^*_2}$ (medium leverage) & (c) $\bm{\theta^*_3}$ (low leverage) \\ [7pt]
  \end{tabular}
}
\caption{\double \incolour Flash crash propagation speed (minutes\textsuperscript{-1})through the fund--asset network as a function
of portfolio diversification ($\rho$) and crowding ($\beta$) parameters. The
analysis is repeated for three leverage regimes denoted $\{\bm{\theta^*_1}, \bm{\theta^*_2}, \bm{\theta^*_3}\}$
in decreasing order of magnitude.}
\label{fig:crowding_speed}
\end{figure}

Figure \ref{fig:crowding_speed} presents the speed of flash crash propagation as a function
of fund and asset diversification in the same three leverage regimes described in the previous section. As mentioned previously, and discussed in detail in  Section \ref{ssub:flash_crash_propagation_speed},
the speed we report here is inversely related to the time taken for 80\% of subsequent
flash crashes to occur, following the observation of the first flash crash.

Figure \ref{fig:crowding_speed}(a) demonstrates that at high leverage, portfolio crowding makes little difference to the speed at which flash crashes propagate between assets. Instead, the speed strongly depends on network diversification ($\beta$, in our notation). The previous section established that under high leverage, enhanced diversification is not beneficial to systemic stability. Figure \ref{fig:crowding_speed}(a) shows that enhanced diversification leads to faster flash crash propagation between assets. The apparent stimulating effect of diversification on propagation
speed may indicate that if a set of deleveraging funds wields enough market impact
to cause a flash crash in one asset, it likely also wields enough market impact
to cause a flash crash in many assets (it is unlikely to have exactly the
right level of impact needed to distress one and only one asset) and so enhanced
diversification can simply become an enhanced channel for distress propagation. Enhanced diversification also implies that, conditional on entering distress, a fund will send sell orders to a larger fraction of assets in the network. Under high leverage, the resulting price impact causes these assets to undergo flash crashes at similar times during the simulated trading session, resulting in a higher speed measurement. The observed maximum speed of approximately 200 flash crashes per minute corresponds to a time difference of just 5 minutes between the first flash crash and the 80th percentile flash crash. Such a short window places strict constraints on any \emph{ex post} regulatory interventions, and instead indicates that \emph{ex ante} measures may be more useful as policy tools.

Figure \ref{fig:crowding_speed}(b) presents flash crash propagation speed results in a moderate leverage regime. Here we find that speed remains low regardless of diversification for non-crowded portfolios. In the previous section we observed that default cascades were unlikely for moderate to high diversification for moderate leverage and low crowding (Figure \ref{fig:crowding_contagion}(c)). However, cascades were observed to occur for low diversification in this regime. The low speed at which flash crashes propagate from asset to asset in this case may be due to the fact that the fund-asset network is only sparsely connected, meaning that the effective network \emph{diameter} (in a graph-theoretic sense) is higher. A higher diameter implies that some assets are more weakly-connected to the network and hence distress takes longer to propagate to such nodes. The results for high-crowding regimes are similar to those for high-crowding regimes at higher leverage, though this observation only holds for low to moderate diversification $\rho$. When $\rho$ increases beyond approximately 50\% Figure \ref{fig:crowding_speed}(b) shows that increased diversification reduces flash crash contagion speed. We note that when portfolios are maximally diversified, there can be little difference between crowded and non-crowded configurations (we revisit this notion in the next section), and so it is not surprising that the flash crash propagation speed is similar for crowded and non-crowded topologies when $\rho \to 1.0$. However, it is not obvious why speed should peak near $\rho=0.5$. As density increases from $\rho=0$ to $\rho=0.5$ in crowded regimes ($\beta > 0$), the number of assets that have non-negligible levels of investment increases. Crucially, however, at low density the effect of this increase is detrimental to systemic stability (as we saw in the previous section). It is not until density increases further that diversification acts to stabilise lower-leverage systems. It is interesting that the peak of flash crash contagion speed occurs at a higher value of the diversification, $\rho$, than the peak probability of contagion. This could be due to the conditional nature of the speed measurement. Speed is only defined when more than one asset experiences a flash crash, and this is likely highly correlated with cases where default cascades occur. If we only consider cases where default cascades occur, it does not necessarily follow that the speed, under this condition, will be maximised at the same level of diversification as the unconditional systemic stability.

Figure \ref{fig:crowding_speed}(c) presents similar results for a low leverage scenario. Again we see that more crowded systems exhibit higher flash crash propagation speed regardless of portfolio diversification, and that the speed reaches a maximum at an intermediate level of diversification. The figure clearly shows that the maximal propagation speed, for crowded scenarios, occurs at a lower level of diversification coefficient $\rho$ for the low leverage scenario, compared to the moderate leverage scenario. This finding is broadly consistent with Figures \ref{fig:crowding_contagion}(c) and \ref{fig:crowding_contagion}(e) which find that probability of contagion reaches a maximum for lower values of $\rho$ for low leverage scenarios compared to moderate leverage scenarios. Since flash crash contagion speed is only defined for Monte Carlo trials where at least two assets experience flash crashes, Figure \ref{fig:crowding_speed}(c) features incomplete series\footnote{We also only report results for Monte Carlo trials with valid distributions. This means that certain parameter settings where only a single Monte Carlo replicate resulted in a valid speed measurement are suppressed as it is not possible to give confidence intervals in such cases.}.

Taken together, Figures \ref{fig:crowding_speed}(a)-(c) demonstrate that flash crash propagation speed increases as leverage increases. The results also demonstrate that for high-crowding regimes, speed is a non-monotone function of portfolio diversification. The non-monotonicity is related to similar non-monotonicity observed for the probability of contagion in Section \ref{sub:probability_and_extent_of_contagion}, though the maxima occur at different levels of diversification when considering speed as opposed to stability. We have also found that, similar to the systemic stability results, dispersed configurations and configurations with no crowding have the lowest flash crash propagation speed. This means that they are not only more robust to default cascades, but also that in such scenarios regulators have a larger time window in which to deploy \emph{ex post} interventions should a deleveraging event occur.

\subsection{Effect of asset allocation distribution}
\label{sub:effect_of_asset_allocation_distribution}

In Section \ref{sub:probability_and_extent_of_contagion} we found the surprising result that in certain circumstances, increased portfolio crowding is beneficial to systemic stability. We are not aware of previous studies that have encountered this phenomenon. It is not apparent in \citet{caccioli2014stability} or \citet{chen2014asset}. In this section we present evidence in support of the hypothesis that the effect is due to the non-uniformity of fund investments between the assets in their portfolio. We thus establish the importance of considering non-uniform asset allocations in future studies.

With reference to Figure \ref{fig:crowding_contagion}(a), we consider the probability of contagion in the high leverage scenario $\bm{\theta^*_1}$. We observe that a higher crowding parameter $\beta$ results in a reduced probability of contagion for most values of portfolio diversification $\rho$, and that this effect becomes more pronounced as $\rho$ increases. This is surprising because $\rho \to 1$ implies that all funds make investments in all assets. The notion of preferentially attaching to a subset of assets therefore has no meaning when funds are simply forced to invest in every asset. This leads us to hypothesise that the non-uniformity of the size of allocations funds make to each asset must have an effect on stability. This would be a potentially important result because a common assumption in the literature is that funds make equal investments in each of the assets in their portfolio, e.g. \citep{acemoglu2015systemic, caccioli2014stability}.

\begin{figure}
\centering
\makebox[\textwidth][c]
{
  \includegraphics{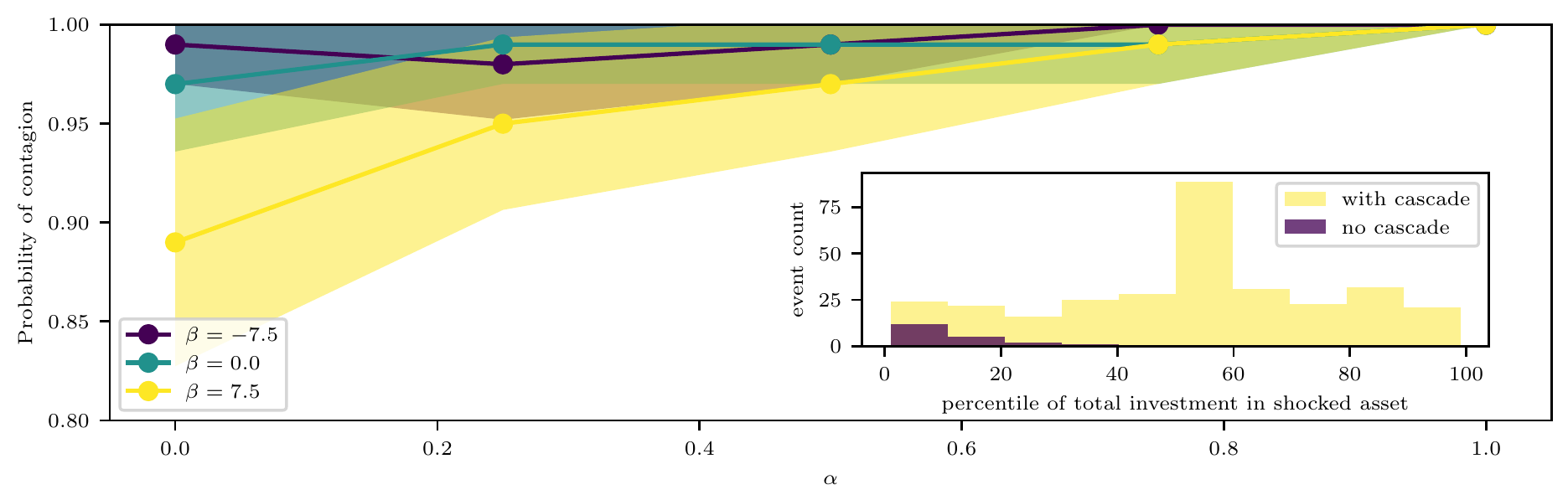}
}
\caption{\double \incolour Probability of contagion with 95\% confidence interval as we interpolate between Gaussian-distributed
and uniformly-distributed asset allocations for each fund. $\alpha \to 1$ implies funds
allocate their capital \emph{equally} to each asset in which they invest.
We fix $\rho=1$ here, such that each fund invests in every asset. This study was
carried out in the high-leverage scenario, $\bm{\theta^*_1}$. Inset: The number
of Monte Carlo trials exhibiting a default cascade as a function of the total level
of investment in the asset that receives the price shock, for the highly-crowded $\beta=7.5$ case.}
\label{fig:alpha_sweep}
\end{figure}

In order to test this hypothesis, we introduce a parameter $\alpha$ that we use to interpolate between funds holding uniformly-allocated portfolios ($\alpha \to 1$) or Gaussian-allocated portfolios ($\alpha \to 0$). All previous results in the present paper set $\alpha=0$. Figure \ref{fig:alpha_sweep} presents the probability of contagion in the high leverage scenario, at maximum diversification ($\rho=1$ such that all funds invest in all assets), as we vary crowding parameter $\beta \in \{ -7.5, 0, 7.5 \}$ and interpolate $\alpha \in \{ 0, 0.25, 0.5, 0.75, 1 \}$. For each $(\alpha, \beta)$ parameter pair we performed 100 Monte-Carlo trials, and report the mean and 95\% confidence interval from the resulting distribution of default cascade occurrence. As hypothesised, we see that the probability of contagion for different levels of crowding converges as we interpolate from Gaussian to uniform portfolios. This confirms that the surprisingly beneficial impact of increased portfolio crowding in high-leverage scenarios is indeed a result of the non-uniformity of asset allocations in our model.

Although we have established a connection between portfolio allocation uniformity and systemic risk, we now consider the reason that this connection exists. We hypothesise that the reduced probability of contagion for non-uniform asset allocations in high-crowding compared to low-crowding scenarios is due to the presence of weakly-connected assets in the fund--asset network.

As described in Section \ref{ssub:fund_asset_bipartite_network}, when funds allocate their investments to assets during network construction, they do so in decreasing investment size order. Since the preferential attachment process selects investment targets without replacement, this means that in crowded scenarios ($\beta > 0$) funds will tend to allocate their largest investment to the same asset as have other funds. This leads to dispersion in the overall amount invested in each asset. As a consequence, some assets receive significantly higher levels of total investment than others. We might expect that assets with lower levels of total investment are, \emph{ceteris paribus}, less systemically important than assets with higher levels of investment. This conjecture must hold in the limit $\sum_{j=1}^{n_f}A_{ij} \to 0$ for some asset $i \in \lbrack1, n_a\rbrack$ since by definition such an asset is disconnected from the network.

The inset graph of Figure \ref{fig:alpha_sweep} presents the number
of Monte Carlo trials exhibiting a default cascade as a function of the total level of investment in the asset that receives the price shock, for the highly-crowded $\beta=7.5$ case, for all values of $\alpha$. We rank assets according to their total level of investment and report the normalised rank (percentile) of the asset that receives the initial shock. The figure clearly shows that when no default cascade occurs (dark shaded bars), the shocked asset did indeed have a low level of total investment compared to other assets. This strongly supports our hypothesis that the surprising benefit of portfolio crowding on systemic stability in high leverage, high diversification scenarios is due to the presence of \emph{weakly-connected} assets. This, in turn, establishes the importance of considering non-uniform asset allocations in future studies of such phenomena. Figure \ref{fig:alpha_sweep} (inset) also demonstrates that the converse does not hold --- default cascades frequently occur even when the shocked asset is weakly-connected. This indicates that the emergent system dynamics are not simply reducible to a statement about the level of investment in a given asset. This finding also has important regulatory consequences. We find that regulators cannot simply ignore the effect of weakly-connected assets when considering systemic stability, as shocks to such assets can readily trigger catastrophic systemic failure.

\FloatBarrier

\section{Conclusion}
\label{sec:discussion_and_further_work}

Our results contribute to the rapidly growing literature on systemic risk. We provide a significant early bridge between the macroscopic domain of networks of shared asset holdings and the microscopic domain of algorithmic trading and flash crashes. This study has shown that the detailed (microscopic) behaviour of automated trading strategies has deep implications for (macroscopic) systemic risk. Our results will be of interest to systemic risk researchers in academia, as well as to regulators who desire enhanced tools to assess the stability of global financial systems.

The present study has been one of the first attempts to augment models of shared asset financial contagion with limit order book driven price impact. Our results lend support to key conclusions from previous macroscopic models, including the destabilising effects of leverage, capital and portfolio crowding. Our model also reproduces expected robust-yet-fragile behaviour where systemic collapse is unlikely, but extensive when it occurs.

Our key findings are:
\begin{enumerate}
    \item Leverage management practice by investment funds engenders a contagion channel for flash crashes to propagate between algorithmically traded assets.
    \item The speed of flash crash propagation strongly depends on the amount of leverage and capital deployed by funds and on network topology, but the dependence is not monotonic in high-crowding scenarios.
    \item Crowding can, under some circumstances, be \emph{beneficial} to systemic stability.
\end{enumerate}

The calibration of the model to realistic behavioural timescales allowed us to move beyond previous macroscopic studies, and to measure the speed at which distress propagates through the financial network during the trading day. We used this approach to characterise the time window available for regulatory interventions, with results suggesting \emph{ex ante} precautions may have higher efficacy than \emph{ex post} reactions.

We found that financial network structure and leverage management policies each have critical impact on both systemic stability and distress propagation speed. Surprisingly, we found that in high-leverage scenarios, under certain conditions, increased portfolio crowding \emph{enhances} systemic stability. We connected this result to the presence of weakly-connected asset nodes in the financial network. Such circumstances do not occur in previous models that conventionally assume uniform portfolio allocations.

Finally, our results provide new insights into the role played by algorithmic trading strategies in the propagation of systemic risk. We find that leverage management policies by funds with overlapping portfolios provide a sufficient contagion channel for the propagation of flash crashes between algorithmically-traded assets. This contagion channel has received little attention in the flash crash simulation literature. Our findings highlight the need for future systemic risk models to move beyond long-established non-crisis price impact assumptions. The path-dependent nature of liquidity provision and flash crashes in asset markets strongly suggests that agent-based methodology is more appropriate than closed-form analysis for future studies.

\subsection*{Future Work}
\label{sub:future_work}

Bringing real-world data into the macroscopic model component constitutes an
important future work direction, particularly in order for recommendations
to be useful to regulators and policy-makers.
The fund-asset network is in
principle available from public 13-f filings (in the USA) though the size of this
dataset would demand careful handling in order to keep simulations tractable\footnote{Although
the simulation has been optimized for a state-of-the-art HPC cluster,
scaling up from tens to tens of thousands of assets represents a significant
challenge.}. Bank-fund margin arrangements are highly proprietary and so it
may be necessary to utilise the work of others such as \citet{ang2011hedge} who
have accessed such datasets.

Further extensions involve the relaxation of the assumptions of the model. It would be straightforward to model non-uniform asset
liquidity and non-uniform leverage and capital across funds. It would be interesting
to see what impact the introduction of such heterogeneity would have on systemic
risk. A particular priority would be to allow funds to enter both long and short
positions. This would then enable us to explore events such as the 2007 \emph{Quant
Meltdown} \citep{khandani2007happened,cont2013running} where it was risk factors, rather than
simply asset holdings, that became crowded. As \citet{capponi2015price} find that
the strategic selection of which assets to sell during distressed deleveraging
has a critical impact on systemic risk, we could extend our work
to explore whether \citet{capponi2015price}'s findings hold when price impact is modelled
realistically.

Exploring strategic decision making by banks in terms of their margin call issuance
could be an interesting future direction, leading naturally into a multi-bank model
in which banks seek to protect their interests. This could include not issuing margin
calls in order to avoid fire sales in assets the bank itself holds. Banks
are also an obvious regulatory target for new policy and interventions.

Finally, liquidity provision by trading agents and, in-particular,
market makers plays a critical role during flash crashes by directly affecting the price impact of distressed sell orders. Our model could be readily extended
to explore the systemic impact of strategic decision making by those agents. In particular, it would be interesting to compare the leverage-induced contagion channel of the present paper with other channels explored in the literature such as arbitrage of exchange traded funds (ETFs) against futures contracts. These contagion channels were each proposed as important by both the \citet{cftc2010findings} in response to the Flash Crash of 6th May 2010 and \citet{jackson2017sterling} in response to the Sterling Flash Event of 7th October 2016. Our model provides a solid foundation for further investigation into the role played by algorithmic trading in causing systemic risk.

\section*{Acknowledgements}
\label{sec:acknowledgements}
Special thanks to Rafael Baptista,
Edith Elkind, and Paul Goldberg.
We would like to acknowledge the use of the University
of Oxford Advanced Research Computing (ARC) facility in carrying out this work
\url{http://dx.doi.org/10.5281/zenodo.22558}.
This work was supported by OxFORD Asset Management (financial support only).

\appendix

\section{Trader Agent Details}
\label{sec:trader_agent_details}

The behavioural details of the trader agent classes are as follows:

\paragraph*{Small Traders}

Small traders place minimally-sized limit orders according to a uniform
distribution around prevailing bid and ask prices on the LOB and thus
are truly zero-intelligence \citep{gode1993allocative}. These
orders may be aggressive and remove liquidity from the book or passive and add
liquidity which may match against subsequent aggressive orders of opposing sign.
Small traders have no dependence on price or volume changes, and we do not model
inventory limits for this agent type. Their slow interaction speed means that
any individual small trader is highly unlikely to build a significant position
during a simulation run. We deviate from \citet{paddrik2012agent}'s implementation
by increasing the spread within which orders can be placed from 10 ticks to 1000
ticks. This change allows our model to better capture the relative price
stylised fact, as discussed in Section \ref{sub:empirical_validation_results}, and is our method of modelling \emph{stub quote} order placement. Stub quotes are orders placed far from prevailing prices that are not intended for execution, but that were identified as playing a role in the Flash Crash of 6th May 2010 \citep{cftc2010findings}.

\paragraph*{Fundamental buyers / sellers}
Fundamental buyers / sellers are in possession of a private valuation for the
asset and will execute when prevailing prices yield a trading surplus \citep{wellman2011trading}.
Fundamental buyers / sellers seek to balance trading impact with immediacy
\citep{paddrik2012agent} and so are represented as gradually building positions
throughout the simulation run. Inventory limits are not modelled for such agents,
however the agents do possess slightly above zero intelligence (known as
$\epsilon$-intelligence). As in \citet{paddrik2012agent}, fundamental buyers /
sellers are sensitive to \emph{toxic order flow} \citep{easley2012flow}. If prices
trend either higher or lower by more than 70 ticks over a one-minute horizon,
fundamental buyers / sellers will avoid placing any orders until the market stabilises.
This behaviour was identified during the real 6th May 2010 Flash Crash by \citet{kirilenko2017flash}.
Fundamental buyers / sellers have an extremely important function in flash crash
generation --- without the action of fundamental buyers reacting to far from fundamental
prices, prices would not rebound following the initial crash.

\paragraph*{Opportunistic Herding}
These agents represent a class of traders that react to external information such as corporate
news dissemination. The order placement direction probability of opportunists
follows a stochastic process. These agents operate over an intermediate
timescale and are modelled with small inventory limits. It is not entirely
clear from \citet{paddrik2012agent} or later related works
\citep{hayes2014agent, paddrik2015effects, bookstaber2015agent} what the impact
of opportunists is on the model dynamics, but since \citet{kirilenko2017flash}
identify this group of real-world market participants we include them for
completeness.

\paragraph*{Market Makers}

Market makers are algorithmic traders that maintain simultaneous limit orders on both sides of the LOB
and exclusively provide (rather than remove) liquidity. As such, they have
been identified as systemically important for the efficient operation of asset
markets \citep{wah2015welfare, bookstaber2015agent}. \citet{paddrik2012agent} model market makers
with small inventory capacity and, furthermore, model the documented tendency for
market makers to retreat from trading by widening spreads or indeed suspending
order placement entirely when they detect toxic order flow \citep{easley2012flow}.
Similarly to fundamental buyers / sellers, when the one-minute price trend exceeds
24 ticks market makers cease trading (calibration based on \citet{kirilenko2017flash}). When the market maker inventory is saturated, the agents attempt to liquidate positions passively. They do this by providing liquidity only on the offsetting side of the LOB.
We model market makers exiting the liquidation state once their absolute inventory reaches half
of its maximum limit.

\paragraph*{High Frequency Traders}

High frequency traders (HFTs) in the model act as very fast market makers first and
foremost, though they also possess a simple shared trading signal
\citep{paddrik2012agent} which skews their order placement side probability
in the direction of LOB imbalances. An imbalance in this context refers to a difference in the quantity of available shares resting on either side of the LOB. The probability that an HFT will place a buy order in the $j$th asset is given by $P(buy)_j=\sfrac{Q_j^{\mathit{bid}}}{(Q_j^{\mathit{bid}}+Q_j^{\mathit{ask}})}$,
where $Q_j^{\mathit{bid}}$ ($Q_j^{\mathit{ask}}$) refers to the total quantity of shares available at the best bid (ask) price \citep{paddrik2012agent}. This signal can be considered a crude prediction algorithm. The intuition behind it is that, when $Q_j^{\mathit{bid}} \gg Q_j^{\mathit{ask}}$ it is likely that the price will rise in the near future, and so it is rational for the HFT agent to buy, in order to profit from the expected price move. The public nature of the signal results
in a well-documented herding effect \citep{menkveld2016economics}.
When their small inventory\footnote{The findings of \citet{cftc2010findings} suggests a total inventory across all HFTs of 3000 contracts. \citet{paddrik2012agent} utilise 3000 contracts per HFT, which remains uncorrected in later papers \citep{hayes2014agent,paddrik2015effects}.} is saturated, HFTs aggressively reduce positions by
taking liquidity. This effect has been implicated in deepening the 6th May
2010 Flash Crash as HFTs passed positions back and forth between themselves
in a ``hot potato'' fashion \citep{abrol2016high, aldrich2016flash}.

\section{Bipartite Network Generation}
\label{sec:bipartite}

Algorithm \ref{algo:random_bipartite} presents our method for constructing random bipartite networks with controllable diversification and crowding.

\begin{algorithm}
\caption{Random Bipartite Networks with Preferential Attachment}
\label{algo:random_bipartite}
\begin{algorithmic}[1]
\REQUIRE $n_f, n_a, \rho, \beta, C^0, \lambda^0, \sigma$
\STATE $A_{ij} \gets 0, \forall i \in \{1, ..., n_f\}, \forall j \in \{1, ..., n_a\}$
\FORALL{$i \in \{1, ..., n_f\}$}

  \STATE \COMMENT{Calculate total value already invested in each asset}
  \FORALL{$j \in \{1, ..., n_a\}$}
    \STATE $total_j \gets \sum_i^{n_f}A_{ij}$
  \ENDFOR

  \STATE \COMMENT{Calculate preferential probability of selecting each asset}
  \FORALL{$j \in \{1, ..., n_a\}$}
    \STATE $r_j \gets rank(j, total)$
    \IF{$\beta<0$}
      \STATE $r_j \gets n_a - r_j + 1$
    \ENDIF
    \STATE $p_j \gets r_j^\beta / \sum_j^{n_a}r_j^\beta$
  \ENDFOR

  \STATE \COMMENT{Calculate fund fractional investment to assets}
  \STATE $k_{\mathit{fund}} \gets max(\mathcal{N}(\rho, \sigma) \times n_a, 1)$
  \FORALL{$k \in \{1, ..., k_{\mathit{fund}}\}$}
    \STATE $j' \gets weightedChoice(n_a, p, \FALSE)$
    \STATE $A_{ij'} \gets nextInvestmentSize(i)$
  \ENDFOR

  \STATE \COMMENT{Convert normalised fractional investments to currency values}
  \STATE $A_{ij} \gets C^0(1+\lambda^0)A_{ij}/\sum_j^{n_a}A_{ij}$
\ENDFOR
\RETURN $A$

\end{algorithmic}
\end{algorithm}

Algorithm \ref{algo:random_bipartite} makes the following function calls:

\paragraph*{$\bm{\mathit{rank}(j, L)}$} returns the index that the $j$th element of list $L$ would receive, if the elements of $L$ were sorted in ascending numerical order.

\paragraph*{$\bm{\mathit{weightedChoice}(n_a, \mathit{weights}, \mathit{replacement})}$} returns $j' \in [1,n_a]$ with probability given by the $j'$th element of the vector $weights$, with or without replacement.

\paragraph*{$\bm{\mathit{nextInvestmentSize}(i)}$} returns values in units of currency representing the sizes of investment fund $i$ makes in each of the $k_{\mathit{fund}}$ assets in its portfolio. Investments are returned in decreasing size order based on repeated samples from a Gaussian distribution $\mathcal{N}(0,1)$ constrained such that the fractional investments to the $k_{\mathit{fund}}$ assets sum to unity.

\clearpage
\bibliographystyle{elsarticle-harv}
\bibliography{main}{}

\begin{thebibliography}{78}
\expandafter\ifx\csname natexlab\endcsname\relax\def\natexlab#1{#1}\fi
\expandafter\ifx\csname url\endcsname\relax
  \def\url#1{\texttt{#1}}\fi
\expandafter\ifx\csname urlprefix\endcsname\relax\def\urlprefix{URL }\fi

\bibitem[{Abrol et~al.(2016)Abrol, Chesir, and Mehta}]{abrol2016high}
Abrol, S., Chesir, B., Mehta, N., 2016. High frequency trading and us stock
  market microstructure: A study of interactions between complexities, risks
  and strategies residing in us equity market microstructure. Financial
  Markets, Institutions \& Instruments 25~(2), 107--165.

\bibitem[{Acemoglu et~al.(2015)Acemoglu, Ozdaglar, and
  Tahbaz-Salehi}]{acemoglu2015systemic}
Acemoglu, D., Ozdaglar, A., Tahbaz-Salehi, A., 2015. Systemic risk and
  stability in financial networks. The American Economic Review 105~(2),
  564--608.

\bibitem[{Acharya et~al.(2017)Acharya, Pedersen, Philippon, and
  Richardson}]{acharya2017measuring}
Acharya, V.~V., Pedersen, L.~H., Philippon, T., Richardson, M., 2017. Measuring
  systemic risk. The Review of Financial Studies 30~(1), 2--47.

\bibitem[{Adrian and Shin(2010)}]{adrian2010liquidity}
Adrian, T., Shin, H.~S., 2010. Liquidity and leverage. Journal of financial
  intermediation 19~(3), 418--437.

\bibitem[{Aldrich et~al.(2016)Aldrich, Grundfest, and
  Laughlin}]{aldrich2016flash}
Aldrich, E.~M., Grundfest, J., Laughlin, G., 2016. The flash crash: a new
  deconstruction. Available at SSRN 2721922.
\newline\urlprefix\url{http://dx.doi.org/10.2139/ssrn.2721922}

\bibitem[{Allen and Babus(2008)}]{allen2008networks}
Allen, F., Babus, A., 2008. Networks in finance. Wharton Financial Institutions
  Center Working Paper~(07-08).
\newline\urlprefix\url{http://dx.doi.org/10.2139/ssrn.1094883}

\bibitem[{Ang et~al.(2011)Ang, Gorovyy, and Van~Inwegen}]{ang2011hedge}
Ang, A., Gorovyy, S., Van~Inwegen, G.~B., 2011. Hedge fund leverage. Journal of
  Financial Economics 102~(1), 102--126.

\bibitem[{Arinaminpathy et~al.(2012)Arinaminpathy, Kapadia, and
  May}]{arinaminpathy2012size}
Arinaminpathy, N., Kapadia, S., May, R.~M., 2012. Size and complexity in model
  financial systems. Proceedings of the National Academy of Sciences 109~(45),
  18338--18343.

\bibitem[{Barab{\'a}si and Albert(1999)}]{barabasi1999preferential}
Barab{\'a}si, A.-L., Albert, R., 1999. Emergence of scaling in random networks.
  Science 286~(5439), 509--512.

\bibitem[{Battiston et~al.(2016)Battiston, Caldarelli, D'Errico, and
  Gurciullo}]{battiston2016leveraging}
Battiston, S., Caldarelli, G., D'Errico, M., Gurciullo, S., 2016. Leveraging
  the network: a stress-test framework based on debtrank. Available at SSRN
  2571218.

\bibitem[{Benoit et~al.(2017)Benoit, Colliard, Hurlin, and
  P{\'e}rignon}]{benoit2016risks}
Benoit, S., Colliard, J.-E., Hurlin, C., P{\'e}rignon, C., 2017. Where the
  risks lie: A survey on systemic risk. Review of Finance 21~(1), 109--152.

\bibitem[{Bisias et~al.(2012)Bisias, Flood, Lo, and
  Valavanis}]{bisias2012survey}
Bisias, D., Flood, M.~D., Lo, A.~W., Valavanis, S., 2012. A survey of systemic
  risk analytics. US Department of Treasury, Office of Financial
  Research~(0001).

\bibitem[{Bookstaber and Paddrik(2015)}]{bookstaber2015agent}
Bookstaber, R., Paddrik, M.~E., 2015. An agent-based model for crisis liquidity
  dynamics. OFR Working Paper 15~(18).

\bibitem[{Booth(2016)}]{booth2016algorithmic}
Booth, A., April 2016. Automated algorithmic trading: machine learning and
  agent-based modelling in complex adaptive financial markets. Ph.D. thesis,
  University of Southampton.
\newline\urlprefix\url{https://eprints.soton.ac.uk/397453/}

\bibitem[{Brunnermeier and Pedersen(2009)}]{brunnermeier2009market}
Brunnermeier, M.~K., Pedersen, L.~H., 2009. Market liquidity and funding
  liquidity. Review of Financial studies 22~(6), 2201--2238.

\bibitem[{Budish et~al.(2015)Budish, Cramton, and Shim}]{budish2015high}
Budish, E.~B., Cramton, P., Shim, J.~J., 2015. The high-frequency trading arms
  race: Frequent batch auctions as a market design response. Chicago Booth
  Research Paper 130~(14-03).

\bibitem[{Caccioli et~al.(2015)Caccioli, Farmer, Foti, and
  Rockmore}]{caccioli2015overlapping}
Caccioli, F., Farmer, J.~D., Foti, N., Rockmore, D., 2015. Overlapping
  portfolios, contagion, and financial stability. Journal of Economic Dynamics
  and Control 51, 50--63.

\bibitem[{Caccioli et~al.(2014)Caccioli, Shrestha, Moore, and
  Farmer}]{caccioli2014stability}
Caccioli, F., Shrestha, M., Moore, C., Farmer, J.~D., 2014. Stability analysis
  of financial contagion due to overlapping portfolios. Journal of Banking \&
  Finance 46, 233--245.

\bibitem[{Capponi and Larsson(2015)}]{capponi2015price}
Capponi, A., Larsson, M., 2015. Price contagion through balance sheet linkages.
  Review of Asset Pricing Studies, rav006.

\bibitem[{Cespa and Foucault(2014)}]{cespa2014illiquidity}
Cespa, G., Foucault, T., 2014. Illiquidity contagion and liquidity crashes.
  Review of Financial Studies 27~(6), 1615--1660.

\bibitem[{CFTC and SEC(2010)}]{cftc2010findings}
CFTC, S., SEC, U., 2010. Findings regarding the market events of may 6, 2010.
  Report of the Staffs of the CFTC and SEC to the Joint Advisory Committee on
  Emerging Regulatory Issues.

\bibitem[{Chen et~al.(2014)Chen, Iyengar, and Moallemi}]{chen2014asset}
Chen, C., Iyengar, G., Moallemi, C.~C., 2014. Asset-based contagion models for
  systemic risk. working paper.

\bibitem[{Chen et~al.(2012)Chen, Chang, and Du}]{chen2012agent}
Chen, S.-H., Chang, C.-L., Du, Y.-R., 2012. Agent-based economic models and
  econometrics. The Knowledge Engineering Review 27~(02), 187--219.

\bibitem[{Cont(2001)}]{cont2001empirical}
Cont, R., 2001. Empirical properties of asset returns: stylized facts and
  statistical issues. Quantitative Finance 1, 223--236.
\newline\urlprefix\url{https://doi.org/10.1080/713665670}

\bibitem[{Cont and Schaanning(2017)}]{cont2017fire}
Cont, R., Schaanning, E.~F., 2017. Fire sales, indirect contagion and systemic
  stress testing. Working paper, Norges Bank.

\bibitem[{Cont and Wagalath(2013)}]{cont2013running}
Cont, R., Wagalath, L., 2013. Running for the exit: distressed selling and
  endogenous correlation in financial markets. Mathematical Finance 23~(4),
  718--741.

\bibitem[{Cristelli et~al.(2010)Cristelli, Alfi, Pietronero, and
  Zaccaria}]{cristelli2010liquidity}
Cristelli, M., Alfi, V., Pietronero, L., Zaccaria, A., 2010. Liquidity crisis,
  granularity of the order book and price fluctuations. The European Physical
  Journal B-Condensed Matter and Complex Systems 73~(1), 41--49.

\bibitem[{Dechow et~al.(2001)Dechow, Hutton, Meulbroek, and
  Sloan}]{dechow2001shortsqueeze}
Dechow, P.~M., Hutton, A.~P., Meulbroek, L., Sloan, R.~G., 2001. Short-sellers,
  fundamental analysis, and stock returns. Journal of Financial Economics
  61~(1), 77 -- 106.
\newline\urlprefix\url{http://www.sciencedirect.com/science/article/pii/S0304405X01000563}

\bibitem[{Di~Gangi et~al.(2018)Di~Gangi, Lillo, and Pirino}]{di2015assessing}
Di~Gangi, D., Lillo, F., Pirino, D., 2018. Assessing systemic risk due to fire
  sales spillover through maximum entropy network reconstruction. Available at
  SSRN 2639178.
\newline\urlprefix\url{http://dx.doi.org/10.2139/ssrn.2639178}

\bibitem[{Easley et~al.(2012)Easley, de~Prado, and O'Hara}]{easley2012flow}
Easley, D., de~Prado, M. M.~L., O'Hara, M., 2012. Flow toxicity and liquidity
  in a high-frequency world. Review of Financial Studies 25~(5), 1457--1493.

\bibitem[{Fagiolo et~al.(2017)Fagiolo, Guerini, Lamperti, Moneta, and
  Roventini}]{fagiolo2017validation}
Fagiolo, G., Guerini, M., Lamperti, F., Moneta, A., Roventini, A., Sep 2017.
  Validation of agent-based models in economics and finance. LEM Papers Series
  2017/23, Laboratory of Economics and Management (LEM), Sant'Anna School of
  Advanced Studies, Pisa, Italy.

\bibitem[{Fagiolo et~al.(2007)Fagiolo, Moneta, and
  Windrum}]{fagiolo2007critical}
Fagiolo, G., Moneta, A., Windrum, P., 2007. A critical guide to empirical
  validation of agent-based models in economics: methodologies, procedures, and
  open problems. Computational Economics 30~(3), 195--226.

\bibitem[{Fagiolo and Roventini(2017)}]{fagiolo2017}
Fagiolo, G., Roventini, A., 2017. Macroeconomic policy in dsge and agent-based
  models redux: New developments and challenges ahead. Journal of Artificial
  Societies and Social Simulation 20~(1), 1.

\bibitem[{Fama and French(1993)}]{fama93commonrisk}
Fama, E.~F., French, K.~R., 1993. Common risk factors in the returns on stocks
  and bonds. Journal of Financial Economics 33, 3--56.

\bibitem[{Farmer and Foley(2009)}]{farmer2009economy}
Farmer, J.~D., Foley, D., 2009. The economy needs agent-based modelling. Nature
  460~(7256), 685--686.

\bibitem[{Farmer et~al.(2004)Farmer, Gillemot, Lillo, Mike, and
  Sen}]{farmer2004really}
Farmer, J.~D., Gillemot, L., Lillo, F., Mike, S., Sen, A., 2004. What really
  causes large price changes? Quantitative finance 4~(4), 383--397.

\bibitem[{Farmer et~al.(2005)Farmer, Patelli, and Zovko}]{farmer2005predictive}
Farmer, J.~D., Patelli, P., Zovko, I.~I., 2005. The predictive power of zero
  intelligence in financial markets. Proceedings of the National Academy of
  Sciences of the United States of America 102~(6), 2254--2259.

\bibitem[{Franke and Westerhoff(2012)}]{franke2012structural}
Franke, R., Westerhoff, F., 2012. Structural stochastic volatility in asset
  pricing dynamics: Estimation and model contest. Journal of Economic Dynamics
  and Control 36~(8), 1193--1211.

\bibitem[{Gai et~al.(2011)Gai, Haldane, and Kapadia}]{gai2011complexity}
Gai, P., Haldane, A., Kapadia, S., 2011. Complexity, concentration and
  contagion. Journal of Monetary Economics 58~(5), 453--470.

\bibitem[{Gai and Kapadia(2010)}]{gai2010contagion}
Gai, P., Kapadia, S., 2010. Contagion in financial networks. In: Proceedings of
  the Royal Society of London A: Mathematical, Physical and Engineering
  Sciences. The Royal Society, p. rspa20090410.

\bibitem[{Gerig(2015)}]{gerig2015high}
Gerig, A., 2015. High-frequency trading synchronizes prices in financial
  markets. Available at SSRN 2173247.
\newline\urlprefix\url{http://dx.doi.org/10.2139/ssrn.2173247}

\bibitem[{Gode and Sunder(1993)}]{gode1993allocative}
Gode, D.~K., Sunder, S., 1993. Allocative efficiency of markets with
  zero-intelligence traders: Market as a partial substitute for individual
  rationality. Journal of political economy, 119--137.

\bibitem[{Golub et~al.(2012)Golub, Keane, and Poon}]{golub2012high}
Golub, A., Keane, J., Poon, S.-H., 2012. High frequency trading and mini flash
  crashes. arXiv preprint arXiv:1211.6667.

\bibitem[{Gould et~al.(2013)Gould, Porter, Williams, McDonald, Fenn, and
  Howison}]{gould2013lobs}
Gould, M.~D., Porter, M.~A., Williams, S., McDonald, M., Fenn, D.~J., Howison,
  S.~D., 2013. Limit order books. Quantitative Finance 13~(11), 1709--1742.

\bibitem[{Greenwood et~al.(2015)Greenwood, Landier, and
  Thesmar}]{greenwood2015vulnerable}
Greenwood, R., Landier, A., Thesmar, D., 2015. Vulnerable banks. Journal of
  Financial Economics 115~(3), 471--485.

\bibitem[{Haldane and May(2011)}]{haldane2011systemic}
Haldane, A.~G., May, R.~M., 2011. Systemic risk in banking ecosystems. Nature
  469~(7330), 351--355.

\bibitem[{Haldane and Turrell(2018)}]{haldane2018interdisciplinary}
Haldane, A.~G., Turrell, A.~E., 2018. An interdisciplinary model for
  macroeconomics. Oxford Review of Economic Policy 34~(1-2), 219--251.

\bibitem[{Hanson(2011)}]{hanson2011effects}
Hanson, T.~A., 2011. The effects of high frequency traders in a simulated
  market. In: Midwest Finance Association 2012 Annual Meetings Paper.

\bibitem[{Hayes et~al.(2014)Hayes, Todd, Chaidarun, Tepsuporn, Beling, and
  Scherer}]{hayes2014agent}
Hayes, R., Todd, A., Chaidarun, N., Tepsuporn, S., Beling, P., Scherer, W.,
  2014. An agent-based financial simulation for use by researchers. In:
  Proceedings of the 2014 Winter Simulation Conference. IEEE Press, pp.
  300--309.

\bibitem[{Huang et~al.(2013)Huang, Vodenska, Havlin, and
  Stanley}]{huang2013cascading}
Huang, X., Vodenska, I., Havlin, S., Stanley, H.~E., 2013. Cascading failures
  in bi-partite graphs: model for systemic risk propagation. Scientific reports
  3, 1219.

\bibitem[{Jackson et~al.(2017)Jackson, Crowley-Reidy, and
  Schrimpf}]{jackson2017sterling}
Jackson, R., Crowley-Reidy, L., Schrimpf, A., 2017. The sterling `flash event'
  of 7 october 2016.

\bibitem[{Jacob~Leal and Napoletano(2017)}]{jacob2017market}
Jacob~Leal, S., Napoletano, M., 2017. Market stability vs. market resilience:
  Regulatory policies experiments in an agent-based model with low- and
  high-frequency trading. Journal of Economic Behavior \& Organization, --.

\bibitem[{Jacob~Leal et~al.(2016)Jacob~Leal, Napoletano, Roventini, and
  Fagiolo}]{jacob2016rock}
Jacob~Leal, S., Napoletano, M., Roventini, A., Fagiolo, G., 2016. Rock around
  the clock: An agent-based model of low- and high-frequency trading. Journal
  of Evolutionary Economics 26~(1), 49--76.

\bibitem[{Johnson et~al.(2012)Johnson, Zhao, Hunsader, Meng, Ravindar, Carran,
  and Tivnan}]{johnson2012financial}
Johnson, N., Zhao, G., Hunsader, E., Meng, J., Ravindar, A., Carran, S.,
  Tivnan, B., 2012. Financial black swans driven by ultrafast machine ecology.
  Available at SSRN 2003874.

\bibitem[{Khandani and Lo(2007)}]{khandani2007happened}
Khandani, A., Lo, A., 2007. What happened to the quants in august 2007?(digest
  summary). Journal of investment management 5~(4), 29--78.

\bibitem[{Kirilenko et~al.(2017)Kirilenko, Kyle, Samadi, and
  Tuzun}]{kirilenko2017flash}
Kirilenko, A., Kyle, A.~S., Samadi, M., Tuzun, T., 2017. The flash crash:
  High-frequency trading in an electronic market. The Journal of Finance
  72~(3), 967--998.
\newline\urlprefix\url{http://dx.doi.org/10.1111/jofi.12498}

\bibitem[{Kirman and Teyssiere(2002)}]{kirman2002microeconomic}
Kirman, A., Teyssiere, G., 2002. Microeconomic models for long memory in the
  volatility of financial time series. Studies in Nonlinear Dynamics \&
  Econometrics 5~(4).

\bibitem[{Krapivsky et~al.(2000)Krapivsky, Redner, and
  Leyvraz}]{krapivsky2000connectivity}
Krapivsky, P.~L., Redner, S., Leyvraz, F., Nov 2000. Connectivity of growing
  random networks. Phys. Rev. Lett. 85, 4629--4632.

\bibitem[{LeBaron(2000)}]{lebaron2000agent}
LeBaron, B., 2000. Agent-based computational finance: Suggested readings and
  early research. Journal of Economic Dynamics and Control 24~(5), 679--702.

\bibitem[{Menkveld(2015)}]{menkveld2015crowded}
Menkveld, A.~J., 2015. Crowded trades: an overlooked systemic risk for central
  clearing counterparties. In: AFA 2015 Boston Meetings.

\bibitem[{Menkveld(2016)}]{menkveld2016economics}
Menkveld, A.~J., 2016. The economics of high-frequency trading: Taking stock.
  Annual Review of Financial Economics 8, 1--24.

\bibitem[{Newman(2003)}]{newman2003structure}
Newman, M.~E., 2003. The structure and function of complex networks. SIAM
  review 45~(2), 167--256.

\bibitem[{Nier et~al.(2007)Nier, Yang, Yorulmazer, and
  Alentorn}]{nier2007network}
Nier, E., Yang, J., Yorulmazer, T., Alentorn, A., 2007. Network models and
  financial stability. Journal of Economic Dynamics and Control 31~(6),
  2033--2060.

\bibitem[{Paddrik et~al.(2012)Paddrik, Hayes, Todd, Yang, Beling, and
  Scherer}]{paddrik2012agent}
Paddrik, M., Hayes, R., Todd, A., Yang, S., Beling, P., Scherer, W., 2012. An
  agent based model of the e-mini {S\&P} 500 applied to flash crash analysis.
  In: 2012 IEEE Conference on Computational Intelligence for Financial
  Engineering \& Economics (CIFEr). IEEE, pp. 1--8.

\bibitem[{Paddrik et~al.(2015)Paddrik, Hayes, Beling, and
  Scherer}]{paddrik2015effects}
Paddrik, M.~E., Hayes, R., Beling, P., Scherer, W., 2015. Effects of limit
  order book information level on market stability metrics. Journal of Economic
  Interaction and Coordination~(forthcoming).

\bibitem[{Panayi et~al.(2012)Panayi, Harman, and Wetherilt}]{panayi2012agent}
Panayi, E., Harman, M., Wetherilt, A., 2012. Agent-based modelling of stock
  markets using existing order book data. In: International Workshop on
  Multi-Agent Systems and Agent-Based Simulation. Springer, pp. 101--114.

\bibitem[{Serri et~al.(2016)Serri, Caldarelli, and Cimini}]{serri2016interbank}
Serri, M., Caldarelli, G., Cimini, G., 2016. How the interbank market becomes
  systemically dangerous: an agent-based network model of financial distress
  propagation. arXiv preprint arXiv:1611.04311.

\bibitem[{Shleifer and Vishny(2011)}]{shleifer2011fire}
Shleifer, A., Vishny, R., 2011. Fire sales in finance and macroeconomics. The
  Journal of Economic Perspectives 25~(1), 29--48.

\bibitem[{Sias et~al.(2016)Sias, Turtle, and Zykaj}]{sias2016hedge}
Sias, R., Turtle, H.~J., Zykaj, B., 2016. Hedge fund crowds and mispricing.
  Management Science 62~(3), 764--784.

\bibitem[{Tesfatsion(2002)}]{tesfatsion2002agent}
Tesfatsion, L., 2002. Agent-based computational economics: Growing economies
  from the bottom up. Artificial life 8~(1), 55--82.

\bibitem[{Torii et~al.(2015)Torii, Izumi, and Yamada}]{torii2015shock}
Torii, T., Izumi, K., Yamada, K., 2015. Shock transfer by arbitrage trading:
  analysis using multi-asset artificial market. Evolutionary and Institutional
  Economics Review 12~(2), 395--412.

\bibitem[{T{\'o}th et~al.(2011)T{\'o}th, Lemperiere, Deremble, De~Lataillade,
  Kockelkoren, and Bouchaud}]{toth2011anomalous}
T{\'o}th, B., Lemperiere, Y., Deremble, C., De~Lataillade, J., Kockelkoren, J.,
  Bouchaud, J.-P., 2011. Anomalous price impact and the critical nature of
  liquidity in financial markets. Physical Review X 1~(2), 021006.

\bibitem[{Vuorenmaa and Wang(2014)}]{vuorenmaa2014agent}
Vuorenmaa, T.~A., Wang, L., 2014. An agent-based model of the flash crash of
  may 6, 2010, with policy implications. Available at SSRN 2336772.

\bibitem[{Wah and Wellman(2015)}]{wah2015welfare}
Wah, E., Wellman, M.~P., 2015. Welfare effects of market making in continuous
  double auctions. In: Proceedings of the 2015 International Conference on
  Autonomous Agents and Multiagent Systems. International Foundation for
  Autonomous Agents and Multiagent Systems, pp. 57--66.

\bibitem[{Wellman(2011)}]{wellman2011trading}
Wellman, M.~P., 2011. Trading agents. Synthesis Lectures on Artificial
  Intelligence and Machine Learning 5~(3), 1--107.

\bibitem[{Wooldridge(2001)}]{wooldridge2001multi}
Wooldridge, M.~J., 2001. Multi-agent systems: an introduction. Wiley
  Chichester.

\bibitem[{Zarinelli et~al.(2015)Zarinelli, Treccani, Farmer, and
  Lillo}]{zarinelli2015beyond}
Zarinelli, E., Treccani, M., Farmer, J.~D., Lillo, F., 2015. Beyond the square
  root: Evidence for logarithmic dependence of market impact on size and
  participation rate. Market Microstructure and Liquidity 1~(02), 1550004.

\bibitem[{Zovko et~al.(2002)Zovko, Farmer, et~al.}]{zovko2002power}
Zovko, I., Farmer, J.~D., et~al., 2002. The power of patience: a behavioural
  regularity in limit-order placement. Quantitative Finance 2~(5), 387--392.

\end{thebibliography}

\end{document}